\documentclass[a4paper,11pt]{article}
\usepackage{jheppub} 
\usepackage{lineno}
\usepackage{subcaption}
 \usepackage{graphics}
 \usepackage{tcolorbox}
 \usepackage{float}
 \usepackage{tikz-cd} 
\usepackage{url}

\usepackage{float}
\usepackage{standalone}
\usepackage{bm}
\usepackage{cleveref}
\usepackage{subcaption}
\usepackage{graphicx} 
\usepackage{array}
\usepackage{mathtools}
\usepackage{paralist}
\usepackage{amsmath}
\usepackage{physics}
\usepackage{float}
\usepackage{xcolor}
\usepackage{amsfonts}
\usepackage{amssymb}
\usepackage{verbatim}
 \usepackage{indentfirst}
\usepackage{tikz}
\usepackage{soul}
\usepackage{pifont}
\usepackage{forest}
\usepackage{tikz-cd}
\tikzcdset{
	arrow style=tikz,
	diagrams={>={Straight Barb[scale=0.6]}},
    every arrow/.append style={line width=0.4pt}
}
\usetikzlibrary{calc}
\tikzset{curve/.style={settings={#1},to path={(\tikztostart)
    .. controls ($(\tikztostart)!\pv{pos}!(\tikztotarget)!\pv{height}!270:(\tikztotarget)$)
    and ($(\tikztostart)!1-\pv{pos}!(\tikztotarget)!\pv{height}!270:(\tikztotarget)$)
    .. (\tikztotarget)\tikztonodes}},
    settings/.code={\tikzset{quiver/.cd,#1}
        \def\pv##1{\pgfkeysvalueof{/tikz/quiver/##1}}},
    quiver/.cd,pos/.initial=0.25,height/.initial=0}
\tikzset{
	solid node/.style={circle,draw,inner sep=1.2,fill=black},
	hollow node/.style={circle,draw,inner sep=1.2},
	left label/.style={above left,midway},
	right label/.style={above right,midway}
}
\usepackage{caption, subcaption}
 \usepackage{tikz}
\usepackage{multirow} 
\usepackage{makecell}
\usepackage{longtable}

\newtheorem{thm}{Theorem}[section]

\newtheorem{conj}[thm]{Conjecture}

\newtheorem{result}[thm]{Claim}
\newtheorem{prob}[thm]{Problem}

\def\C{\mathbb{C}}

\def\Z{\mathbb{Z}}

\def\b{\beta}

\def\g{\gamma}

\def\r{\rho}

\def\t{\tau}

\def\w{\omega}

\def\z{\zeta}
\def\D{\Delta}
\def\G{\Gamma}

\def\P{\Psi}

\def\beq{\begin{eqnarray}}
\def\eeq{\end{eqnarray}}
\def\beq#1\eeq{\begin{align}#1\end{align}}

\title{On holographic duals of certain isolated weighted homogeneous Gorenstein cDV singularities.}
\author[a]{Yuanyuan Fang}
\author[b]{, Zekai Yu}
\affiliation[a]{Dept.\ of Mathematics, Tsinghua University, Beijing, 10084, China}
\affiliation[b]{Qiuzhen College, Tsinghua University, Beijing, 10084, China}
\emailAdd{fangyy21@mails.tsinghua.edu.cn}
\emailAdd{yuzk23@mails.tsinghua.edu.cn}
\abstract{We employ a novel approach, based on homological mirror symmetry for Landau–Ginzburg models, to demonstrate the non-existence of crepant resolutions for certain weighted homogeneous Gorenstein compound Du Val singularities. Physically, this implies that such singularities cannot serve as holographic backgrounds for four-dimensional $\mathcal N=1$ superconformal quiver gauge theories realized on the worldvolume of a large number of D3-branes placed at the singular locus. This is confirmed by enumerating all consistent quiver gauge theories.
}

\begin{document}

\maketitle
\newpage

\section{Introduction}

Compound Du Val (cDV) singularities form an important class of three-dimensional geometric singularities. They have a concrete description as hypersurfaces in $\C^4$, which enables explicit computations. Moreover, they are closely related to the canonical $ADE$ surface singularities, appearing as one-parameter deformations thereof. Among them, the \textit{weighted homogeneous} cDV singularities are distinguished by the presence of an evident dilational $\C^\ast$-action. A class of these singularities is displayed in Table \ref{table:cdvlist} \footnote{More precisely, the hypersurface equations in the list are defined only up to what is called a \textit{weight-one deformations}, i.e., deformations that preserve the original $\C^\ast$-actions.}. The meaning of the rightmost column will be explained later. 
\begin{table}[H]
\begin{centering}
\resizebox{0.8\columnwidth}{!}{
	\begin{tabular}{ |c|c|c|c|}
		\hline
		$J$ &  singularity  &  $K$-stable range of parameters   \\ \hline
		$A_{N-1}$ &$x_1^2+x_2^2+x_3^N+z^k=0$ &$\frac{N}{2}<k<2N, N\geq 2$ \\ \hline
  
		$~$   & $x_1^2+x_2^2+x_3^N+x_3 z^k=0$&$\frac{N^2-1}{2N-1}<k<2N-2,N\geq 2$\\ \hline
		
		$D_N$   & $x_1^2+x_2^{N-1}+x_2x_3^2+z^k=0$ &$\frac{2 N^2-8 N+6}{2 N-3}<k<4 N-4, N\geq 4$\\    
  \hline
		$~$   &$x_1^2+x_2^{N-1}+x_2x_3^2+z^k x_3=0$ &$\makecell{\left(N=4,5,1<k<2N\right)\\
       \left(N\geq 6,\frac{N^2-4 N}{2 N-2}<k<2 N\right)}$ \\     \hline
		
		$E_6$ & $x_1^2+x_2^3+x_3^4+z^k=0$  & $1<k<24$ \\     \hline
		$~$  & $x_1^2+x_2^3+x_3^4+z^k x_3=0$  &  $1<k<18$ \\     \hline
		$~$   & $x_1^2+x_2^3+x_3^4+z^k x_2=0$  & $1<k<16$ \\     \hline
		
		$E_7$  & $x_1^2+x_2^3+x_2x_3^3+z^k=0$   & $1<k<36$ \\     \hline
		$~$   & $x_1^2+x_2^3+x_2x_3^3+z^kx_3=0$   &$1<k<28$  \\     \hline

		$E_8$   & $x_1^2+x_2^3+x_3^5+z^k=0$   &$1<k<60$  \\     \hline
		$~$    & $x_1^2+x_2^3+x_3^5+z^k x_3=0$ &$2<k<48$  \\     \hline
		$~$    & $x_1^2+x_2^3+x_3^5+z^k x_2=0$ &$1<k<40$ \\     \hline
	\end{tabular}}
\caption{Three-fold isolated weighted homogeneous compound Du Val singularities in \cite{Wang_2016,xie20154dn2scftsingularity}. The rightmost row shows the $K$-stable range of parameters.}
\label{table:cdvlist}
\end{centering}
\end{table}
The additional symmetry makes them particularly useful in string-theoretic applications. For example, when the singularities are isolated, they can serve as input data in the type IIB geometric engineering program. The outcomes of geometric engineering are four‐dimensional $\mathcal N=2$ Argyres–Douglas superconformal field theories (SCFTs) \cite{Wang_2016,xie20154dn2scftsingularity}. This provides an alternative geometric realization of certain Argyres–Douglas theories that are more conventionally obtained via compactification from six dimensions. Isolated cDV singularities also play an important role in M‐theory geometric engineering \cite{Intriligator_1997,Apruzzi_2020,Closset_2022,De_Marco_2022}, where they give rise to five-dimensional SCFTs of rank zero.

Three‐fold singularities $X$ furthermore appear in the context of AdS/CFT correspondence. In the large-$N$ limit, four‐dimensional 
$\cal N$=1 superconformal field theories on D3 branes is dual to type IIB string theory on $AdS_5\times L_5$.\cite{Maldacena:1997re,Klebanov:1998hh,eager2012superconformal}, where $L_5$ denotes the \textit{link}, a.k.a. the associated \textit{Sasaki-Einstein} manifold, of the corresponding three‐fold singularity $X$. In this holographic framework, one expects the coordinate ring of the singularity to correspond - at least heuristically - to the chiral ring of the dual 
$\cal N$=1 SCFT \cite{Xie:2019qmw,collins2016k}. 

Compared to the classical geometric engineering of four-dimensional $\mathcal N=2$ supersymmetric field theories, this set-up preserves a smaller number of Poincar\'e supercharges. Because of the reduced supersymmetry, the extent to which the singular geometry determines the field–theoretic data remains an open question. A guiding principle is that whenever the singularity admits a non-commutative crepant resolution (NCCR), the dual superconformal field theory is a quiver gauge theory with superpotential, completely determined by the underlying NCCR \cite{Aspinwall_2012}. 

Later developments showed that for a consistent large-$N$ dual $\cal N$=1 SCFT to exist, additional geometric conditions may be required - such as K-stability with respect to a certain conical 
$\mathbb{C}^*$-action \cite{collins2016k, Xie:2019qmw}. Concretely, for weighted homogeneous cDV singularities in Table.\ref{table:cdvlist}, K-stability with respect to the apparent $\mathbb{C}^*$‑action imposes a constraint on the parameter $k$. Accordingly, in the rightmost column, $k$ is restricted to its K-stable range for physical relevance. Nevertheless, K-stability is not technically required when $N$ is small.

However, the following technical questions remain unresolved: under what conditions does a three-fold singularity $X$ admit a four-dimensional $\mathcal N=1$ superconformal theory dual in the holographic sense? Moreover, when is the dual field theory described by a scale-invariant quiver gauge theory? Answers to these questions are not obvious, and, as observed in \cite{Fazzi_2020} it is strongly possible that some singularities do not provide holographic backgrounds for a superconformal quiver gauge theory. From a mathematical perspective, this occurs when the singularity admits no NCCR. For isolated cDVs, the existence of an NCCR is known to be equivalent to the existence of a crepant resolution (CR)\footnote{By crepant resolution of a variety $X$ we mean a birational morphism 
\[f:Y\rightarrow X,\quad f^\ast K_X=K_Y\]
from a smooth variety $Y$ to the singularity $X$. $K_X$ and $K_Y$ are the canonical divisors on $X$ and $Y$, respectively. We emphasize that $Y$ must be smooth. $X$ is required to be a \textit{Gorenstein} singularity for technical reasons. The intuition behind a birational morphism is that $f$ contracts subvarieties of codimension at least one, and is generically an isomorphism.} \cite{van2004non}, which possibly does not exist. 

The physical relevance of crepant resolutions (and their non-commutative counterparts) can be understood as follows. From the viewpoint of the low-energy worldvolume theory, the geometry is effectively smoothed. When a crepant resolution exists, the brane fractionates into components wrapping the irreducible exceptional loci. By standard string theory reasoning, each wrapped brane gives rise to a gauge group factor. This setup does not rely on a large-$N$ limit and was extensively studied in the early 2000s under the name \textit{fractional branes}, see e.g. \cite{franco2006brane,franco2006gauge,Benvenuti:2004dy,berenstein2002seibergdualityquivergauge}. We apologize for not being able to give a more complete list of references.

To the best of our knowledge, a systematic understanding of the existence of crepant resolutions of cDV singularities remains incomplete, particularly for those of compound $E_n$ type ($cE_n$). However for quasi-homogeneous isolated cDV singularities, several physics-based results exist, notably including the examples listed in Table \ref{table:cdvlist}. In \cite{De_Marco_2022}, Valandro and collaborators studied M-theory geometric engineering on these singularities. They analyzed the Higgs branch structure of the resulting five-dimensional SCFTs using a \textit{gauge-theoretic} strategy, i.e., treating the 5d theory as a reduction of a seven-dimensional $\mathcal N=1$ gauge theory with a varying adjoint Higgs field $\Phi$. The seven-dimensional theory itself can be viewed as geometrically engineered by putting M-theory on deformed $ADE$ surface singularities. The geometry of the deformed $ADE$ surfaces is encoded in the vacuum expectation values of $\Phi$. Consequently, the resulting low energy field theory can be analyzed using representation-theoretic techniques. Through this approach, they established the answer for existence of crepant resolutions for the singularities in Table.\ref{table:cdvlist}, summarized in the corresponding tables of their works. Similar strategy appeared in other related studies, e.g. \cite{collinucci2021higgsbranches5drankzero,Collinucci_2021,Collinucci_2022,De_Marco_20221}. Morally, this method aligns with the classical algebro-geometric framework used to study the \textit{simultaneous resolution} of families of deformed $ADE$ singularities \cite{BRIESKORN1966,BRIESKORN1968}. For example, the existence of a small resolution of a cDV singularity is related to the possibility of lifting the classifying map from the base - denoted $\C_w$ in \cite{De_Marco_2022} - from  $\mathfrak h/W$ to $\mathfrak h$. Since $\mathfrak h/W$ retains only the invariant polynomials, there is no canonical lift to the Lie algebra $\mathfrak g$ or its Cartan subalgebra $\mathfrak h$. But when additional choices are introduced, such a lift becomes possible. This is the abstract role played by $\Phi$; see for instance \cite{karmazyn2017lengthclassificationthreefoldflops} \footnote{We thank the anonymous referee for drawing our attention to this reference.}.

In this paper, we propose an alternative approach - grounded in symplectic geometry and mirror symmetry - to address the question of existence of crepant resolutions of the $cE_n$ singularities listed in Table.\ref{table:cdvlist}. Physically, this corresponds to understanding the existence of superconformal quiver gauge theory duals associated with these three-fold singularities. Our aim is to present a logically transparent framework that may offer new perspectives on this class of problems.

Specifically, our methods substantiate the following claim, which is somehow a corollary of the results in \cite{De_Marco_2022}
\begin{tcolorbox}
\begin{result}\label{E678claim}
A singularity with $J=E_n$ in the list above, within the $K$-stable range, admits a crepant resolution if and only if it is, up to weight-one deformations, one of the following four types:
$$ \begin{aligned}
 & x^2+y^3+z^4+w^{12}=0,\\
& x^2+y^3+yz^3+w^{18}=0,\\
& x^2+y^3+z^5+w^{30}=0,\\
& x^2+y^3+yz^3+w^2z=0.
\end{aligned}$$
\end{result} 
\end{tcolorbox}
Note that most of the singularities with $J=E_n$ will be genuinely of $cE_{n}$ type. Their conjectured holographic duals appear in \cite{Gubser:1998ia,fang2023dimensionalquotientsingularity4d,karmazyn2017lengthclassificationthreefoldflops}, where the associated quiver diagrams coincide with the affine Dynkin diagrams of $ADE$ types\footnote{One cannot rule out theories that are conformally dual to them.}. The superpotentials in \cite{fang2023dimensionalquotientsingularity4d} agrees with \cite{karmazyn2017lengthclassificationthreefoldflops} as well after integrating out massive adjoints. In the holographic correspondence, weight‑one deformations in the geometry translates to exactly marginal deformations in the field theory. So these deformations neither alter the quiver gauge description nor affect the underlying noncommutative crepant resolution. 

Our methodology unfolds in three main stages:
\begin{enumerate}
    \item 
\textbf{Reduction to two families.} By surveying existing mathematical results, we restrict our attention to only two families of singularities in Table \ref{table:cdvlist}.
\item
\textbf{Symplectic cohomology criterion.} Invoking an important conjecture established in \cite{evans2022symplecticcohomologycompounddu}, which links the negative‐degree symplectic cohomology of the Milnor fiber to the existence of a crepant resolution for a cDV singularity, we reduce the question of “Does this singularity admit a crepant resolution?” to the symplectic cohomology computation of its Milnor fiber, in negative degrees.
\item
\textbf{Computational implementation via mirror symmetry.} Homological mirror symmetry furnishes a concrete tool for these computations: one calculates the Hochschild cohomology of the category of (maximally-)equivariant matrix factorizations of the mirror singularity. The general setup and examples appear in Section \ref{gen} where relevant background is explained, and the full computation is carried out in Section \ref{hos}.
\end{enumerate}
In Section \ref{phy}, we then verify our mathematical conclusions from the physics perspective by enumerating all consistent
$\mathcal{N}=1$ superconformal quiver gauge theories, confirming that no additional cases arise beyond the four identified in Claim \ref{E678claim}. Finally, Section \ref{con} summarizes our findings and offers concluding remarks.

\section{General strategies} \label{gen}
In this section, we briefly review the key mathematical concepts. For an introduction to K‑stability, we refer the readers to \cite{collins2016k}, with its physical applications deferred to Section \ref{phy}. Essential results from birational geometry are collected in \cite{grassi2018topologicalinvariantsalgebraicthreefolds}.

\subsection{$K$-stability of the singularities}
The first ingredient is a necessary condition for a three‑fold singularity to admit a four‑dimensional $\mathcal N=1$ superconformal dual  in the large $N$ limit: the existence of a Ricci‑flat conical metric. In the mathematical framework, this metric is guaranteed by K‑stability of the singularity with respect to a specific $\mathbb{C}^*$‑action \cite{Xie:2019qmw}. Concretely, this $\mathbb{C}^*$‑action splits into two commuting parts: a real scaling along the radial direction of the singularity $X$\footnote{Recall that the general philosophy of AdS/CFT from D3-branes is that, in the near horizon geometry, the radial direction of the normal bundle combines with the world-volume of branes to make $\text{AdS}_5$, while the angular direction remains internal. The radial direction in the current setup is just the radial direction of singular 3-folds.} and the so-called \textit{Reeb flow} on the associated Sasaki-Einstein manifold $L$. Under the AdS/CFT correspondence, the field‑theoretic $U(1)_R$ symmetry is identified with the Reeb flow action, normalized so that the holomorphic top form acquires weight two\footnote{Such normalization is due to the fact that the coordinates $x,y,z,w$ are identified as gauge invariant operators on the field theory side via holographic duality. To identify the Reeb vector with the generator of the $U(1)_R$ symmetry in the field theory, we adopt such a normalization.}. We adopt this normalization throughout this paper. 

The Reeb flow lies within the isometry group of the associated Sasaki-Einstein manifold, and these isometries typically originate from the obvious $U(1)$ actions present in the singularity’s defining equation. Each cDV singularity in Table \ref{table:cdvlist} carries one such manifest $U(1)$ symmetry; suspended singularities with form $uv+f(z,w)=0$ ($cA_n$ types) admit an additional $U(1)$ action, which we do not consider here. Accordingly, we take the Reeb vector to be generated by the manifest $U(1)$ symmetry. One then tests K‑stability with respect to this action by the criteria of \cite{collins2016k}, yielding the inequalities in $N$ and $k$ listed in Table \ref{table:cdvlist}. The physical significance of K‑stability is discussed further in Section \ref{phy}.

\subsection{Reduction of the problem}
Our goal is to investigate the technical problem of existence of superconformal quiver gauge theory duals in large $N$ of the weighted homogeneous singularities for $cE_n$ singularities listed in Table.\ref{table:cdvlist}. It is equivalent to the problem of non-commutative crepant resolutions of the singularity (see e.g. \cite{berenstein2002seibergdualityquivergauge,Aspinwall_2012}). 

Unlike $cA_n$ and $cD_n$ singularities, crepant resolutions for the $cE_n$ families remain largely uncharted. Fortunately, several mathematical constraints enable our analysis, as summarized below:

\paragraph{Resolutions via semi-universal unfolding.} 
For hypersurface singularities of the form
\begin{equation}
    F(x,y,z,w)=f_{ADE}(x,y,z)+w^k=0,
\end{equation}
Brieskorn’s criterion (see e.g. \cite{BRIESKORN1968,peters2024isolatedhypersurfacesingularitiesalgebrageometric}) asserts that a crepant resolution exists precisely when $k$ is an integer multiple of the Coxeter number of the corresponding ADE Lie algebra. For instance, $k=2n-2$ for $D_n$, $k=12$ for $E_6$, $k=18$ for $E_7$ and $k=30$ for $E_8$. This result, however, does not seem to immediately extend to more general deformations such as
\begin{equation}
    F=f_{ADE}(x,y,z)+zw=0,
\end{equation}
for which the singularity structure differs substantially from the simple 
$w^k$ suspension.

\paragraph{Topology of the link and number of exceptional curves.} For hypersurface singularities, results are known relating its birational geometric properties and topology of its link structures\footnote{We thank Prof. Dan Xie for drawing our attention in this direction.}. From \cite{Flenner1981,peters2024isolatedhypersurfacesingularitiesalgebrageometric,Caibar1999} one summarizes as follows: If an isolated weighted homogeneous cDV singularity $X$ admits a small resolution (which is a crepant resolution) $\pi:Y\rightarrow X$, then the number of curves that is contracted is precisely given by the number $f(X)$ of mass deformations of the singularity
\[l=f(X),\]
regarded as defining a 4d $\cal N$=2 Argyres-Douglas SCFT. Namely, it is the number of deformations of mass dimension 1, where the mass dimensions $\Delta$ are normalized such that
\[\sum_{i=1}^4\Delta_i-\Delta_F=1.\]
In particular, the number of exceptional curves is the same for any such a small resolution. This result allows us to explicitly compute the number of curves contracted in such a resolution. In particular, the number of mass parameter for the singularities listed in Table \ref{table:cdvlist} were determined in the appendix of \cite{Giacomelli_2018} and are recorded in Table \ref{cdvmass} below.

\begin{table}[!ht]
	\begin{center}
 \resizebox{1.1\linewidth}{!}{
		\begin{tabular}{|c|c|c|c|c|}
			\hline
			 Singularity &  $f(X)$ & Constraints from $K$-stability \\ \hline
			 $x_1^2+x_2^2+x_3^N+z^k=0$&   $g.c.d(N,k)-1$ & $\frac{N}{2}<k<2N, N\geq 2$ \\ \hline
			$x_1^2+x_2^2+x_3^N+x_3 z^k=0$ &  $g.c.d.(N-1,k)$ & $\frac{N^2-1}{2N-1}<k<2N-2,N\geq 2$\\ \hline
			
			$x_1^2+x_2^{N-1}+x_2x_3^2+z^k=0$ & $\frac{g.c.d.(2N-2,k)+2}{2}$ for $\frac{2N-2}{g.c.d.(2N-2,k)}$ odd; 1 for $k$ and $\frac{2N-2}{g.c.d.(2N-2,k)}$ even; 0 for k odd & $\frac{2 N^2-8 N+6}{2 N-3}<k<4 N-4, N\geq 4$ \\     \hline
			$x_1^2+x_2^{N-1}+x_2x_3^2+z^k x_3=0$& $g.c.d.(N,k)$ for $\frac{N}{g.c.d.(N,k)}$ odd; 0 otherwise & $\makecell{\left(N=4,5,1<k<2N\right)\\
       \left(N\geq 6,\frac{N^2-4 N}{2 N-2}<k<2 N\right)}$\\     \hline
			
			 $x_1^2+x_2^3+x_3^4+z^k=0$&6 for $k=0(mod12)$; 2 for $k=3,6,9(mod12)$; 0 for $k\neq0(mod3)$ & $1<k<24$\\     \hline
			 $x_1^2+x_2^3+x_3^4+z^k x_3=0$ & 6 for $k=0(mod9)$; 0 otherwise & $1<k<18$\\     \hline
		$x_1^2+x_2^3+x_3^4+z^k x_2=0$  & 6 for $k=0(mod8)$; 2 for $k=4(mod8)$; 1 for $k\neq0(mod4)$ & $1<k<16$ \\     \hline
			
		 $x_1^2+x_2^3+x_2x_3^3+z^k=0$& 7 for $k=0(mod 18)$; 1 for k even and $k\neq0(mod 18)$; 0 for k odd & $1<k<36$ \\     \hline
		$x_1^2+x_2^3+x_2x_3^3+z^kx_3=0$ & 7 for $k=0(mod14)$; 1 for k even and $k\neq0(mod14)$; 0 for k odd &$1<k<28$  \\     \hline

			$x_1^2+x_2^3+x_3^5+z^k=0$& 8 for $k=0(mod30)$; 0 otherwise &$1<k<60$  \\     \hline
			$x_1^2+x_2^3+x_3^5+z^k x_3=0$  & 8 for $k=0(mod24)$; 0 otherwise & $1<k<48$ \\     \hline
			$x_1^2+x_2^3+x_3^5+z^k x_2=0$ & 8 for $k=0(mod20)$; 0 otherwise & $1<k<40$ \\     \hline
			
		\end{tabular}
  }
	\end{center}
	\caption{Mass parameters of the cDV singularities in questions, along with their range of K-stability.}
	\label{cdvmass}
\end{table}
As we exclude smooth geometry, a necessary condition for existence of a crepant resolution is $l>0$.
As shown in Table \ref{cdvmass}, one observes that $f(X)=N $ holds precisely for the following families of singularities
$$\begin{aligned}
&x_1^2+x_2^2+x_3^{N+1}+z^{N+1}=0,\\
&x_1^2+x_2^{N-1}+x_2x_3^{2}+z^{2N-2}=0,
\end{aligned}$$
and $f(X)=6,7,8$ respectively for the following singularities
$$\begin{aligned}
&x_1^2+x_2^3+x_3^{4}+z^{12}=0,\\
&x_1^2+x_2^3+x_2x_3^{3}+z^{18}=0,\\
&x_1^2+x_2^3+x_3^{5}+z^{30}=0,\end{aligned}$$
up to weight-one deformations. Moreover, $f(X)=1$ for Morrison-Pinkham example $x^2+y^3+y z^3+z t^{2}=0$. These are the correct values of exceptional curves of known singularities admitting crepant resolutions. Other singularities typically admit small values of $f(X)$, as expected.

In summary, the set of $cE_n$ singularities in Table  \ref{table:cdvlist} which 
\begin{enumerate}
    \item do not fit into the criteria of Brieskorn and
    \item admit nontrivial mass deformations and
    \item are not related to the known ones that can be resolved crepantly by weight-one deformations
    \end{enumerate}
    are summarized in the following 
\begin{tcolorbox}
\begin{prob} \label{problem}
How to determine the existence of crepant resolutions of the following singularities?
    $$\begin{aligned} &cE_6: x^2+y^3+z^4+y t^k:  k=2,3,5,6,7,9,10,11,13,14,15 (f=1);\,k=4,12 (f=2). \\
  &cE_7: x^2+y^3+y z^3+z t^{k}: k=4,6,8,10,12,16,18,20,22,24,26 (f=1).\end{aligned}$$
Here we record the corresponding values of $f$ for reference. 
\end{prob}
\end{tcolorbox}

The main result of the paper is to show via a novel method that none of the singularities in Problem \ref{problem} admit a crepant resolution. This means that they do not admit a 4d $\mathcal N=1$ superconformal quiver gauge theory dual.

\subsection{Symplectic cohomology}
In this subsection, we briefly review the notion of symplectic cohomology and explain its implications on the existence of crepant resolutions.
Our discussion follows the treatments in \cite{lekili2021homologicalmirrorsymmetrymilnor, evans2022symplecticcohomologycompounddu, adaloglou2024symplecticcohomologyquasihomogeneouscan} and the overview in \cite{peters2024isolatedhypersurfacesingularitiesalgebrageometric}.

Consider a hypersurface singularity defined by a polynomial equation $$f(x_1,x_2,x_3,x_{4})=0,$$in $\mathbb{C}^4$, with an isolated singular point at the origin, one studies its deformations via the family $$F(x_1,\cdots ,x_{4},t):=f(x_1,\cdots ,x_{4})+t=0,$$ parametrized by $t\in \C$. For sufficiently small and generic $t$, the set
$$ M_X(t):=\{f^{-1}(t)\}\cap B(0,r)\,,$$
inside a ball $B(0,r)\subset\C^{4}$ is smooth, called the \emph{Milnor fiber}. This small deformation does not alter the behavior at infinity, so its boundary is the same as the boundary of the original singularity, namely the link $$L_X:=\{f^{-1}(0)\}\cap S^{7}_r,$$which is a real $5$-dimensional manifold. This is also called the Sasaki-Einstein manifold associated to $X$.

Each Milnor fiber $M_{X}(t)$ of the singularity $X$ is contained in $\C^4$ and hence can be endowed with the canonical symplectic form. This makes the Milnor fiber a symplectic manifold with contact type boundary, together with a complex structure. Standard reasoning in symplectic geometry requires completing the Milnor fiber by attaching a cylindrical end. This allows one to define its \emph{symplectic cohomology} $SH^\ast(M_X(t))$, with integer grading. 

More intuitively, symplectic cohomology counts critical trajectories of certain canonically defined action functional on $M_X$. These include two parts, the constant orbits, i.e. points, and Reeb orbits on the link manifold. The degrees of such orbits are determined by what is called the \textit{Conley–Zehnder index}. A full definition uses a version of Hamiltonian Floer cohomology, which is beyond the scope of this note. It suffices to think of it as counting trajectories (or solitons) connected by gradient flows determined by Floer's equations (or instantons), similar to what was done in Morse cohomology. Curious readers are referred to e.g. \cite{seidel2010biasedviewsymplecticcohomology} for a review.  As alluded to above, symplectic cohomology splits into two parts:
\begin{itemize}
    \item \textbf{Positive symplectic cohomology} $SH_+$ coming from non‑constant periodic Reeb orbits,
    \item \textbf{Negative symplectic cohomology} $SH_-$ coming from constant orbits, i.e., points. 
\end{itemize}

 The constant‑orbit contribution can be reduced to Morse cohomology of $M_X$, which coincides with the usual singular cohomology $H^*(M_X)$. These fit into a long exact sequence
$$ \cdots\rightarrow H^{\ast-1}(M_X)\rightarrow SH_+^\ast(M_X) \rightarrow SH^\ast (M_X)\rightarrow H^\ast(M_X)\rightarrow \cdots \,,$$
It turns out that $SH^*(M_X(t))$ does not depend on the choice of small deformation $t$, so we denote it simply by $SH^*(M_X)$. Moreover, $SH^*(M_X)$ carries the structure of a graded cohomology ring, which is a symplectic invariant of the Milnor fiber \cite{Seidel_2013}.

For isolated Gorenstein cDV singularities, the symplectic cohomology of the Milnor fiber has a very simple form: 
\begin{align}
   \begin{cases}
       \text{rank }SH^3(M_X)&= \mu\,, \\
       SH^k(M_X)&=0 \text{ for }k=2\text{ or }k\geq 4\,. 
   \end{cases} \label{shpositive}
\end{align}
where $\mu$ denotes the \textit{Milnor number} of the singularity. 

Building on explicit computations on weighted homogeneous isolated cDV singularities, an interesting conjecture was brought up in \cite{evans2022symplecticcohomologycompounddu}, which expects that a compound Du Val singularity has crepant resolution with $l$ irreducible exceptional curves if and only if the symplectic cohomology of its Milnor fiber in all negative degrees has rank $l$. It has been proved for all $cA_n$ singularities in \cite{adaloglou2024symplecticcohomologyquasihomogeneouscan}.

In Section \ref{hos}, we will show that among all candidates, only the four cases highlighted in the introduction exhibit symplectic cohomology groups in negative degrees whose ranks match exactly the number of exceptional curves in a crepant resolution. In particular, the numbers are nonzero. Under the assumption that the “only if” direction of the above conjecture holds, this result leads to the main claim of our paper in the introduction.

Relying this argument on a conjecture outright may seem overly ambitious,  nevertheless, in Section \ref{phy} we will present physical arguments that lend it support.

Moreover, direct computation of symplectic cohomology is difficult. The technical tool for this turns out to be homological mirror symmetry: we do computations in \textit{Hochschild cohomology} in the mirror side.

\subsection{Hochschild cohomology: an introduction}

Hochschild cohomology is widely applied in the study of mirror symmetry, as it is a bridge connecting closed string data to the open string ones. The mathematical essence of Hochschild cohomology is that it encodes the deformations of a structure. The following material is standard in the literature.

In its most elementary form, Hochschild cohomology is used to describe the infinitesimal deformation of an associative algebra up to equivalence. In other word it is the tangent space of the "moduli space" of an associative algebra at a specific point. More concretely, an associative algebra $A$ consists of a $\C$ vector space $V$ and associative $\C$-bilinear multiplication $m_0:V\times V\rightarrow V$. A deformation of the algebra $A$ means a deformation of the multiplication from, say, $m_0(x,y)$ to $m(x,y)$. Being an infinitesimal deformation means that one can expand $m$ near $m_0$ and looks at only the first order terms: $m(x,y)=m_0(x,y)+\epsilon (x,y)$. The constraint is associativity of multiplication
\[m(m(x,y),z)=m(x,m(y,z))\,.\]
The first order equation reads 
\[m_0(\epsilon(x,y),z)+\epsilon(m_0(x,y),z)-m_0(x,\epsilon(y,z))-\epsilon(x,m_0(y,z))=0\,.\]
Simplifying notation one may abbreviate $m_0(x,y)$ by $x \cdot y$. We obtain the cocycle condition
\begin{align}
    x\cdot \epsilon(y,z)-\epsilon(x\cdot y,z)+\epsilon(x,y\cdot z)-\epsilon(x,y)\cdot z=0\,.\label{cocyc}
\end{align}
Two multiplications $m$ and $m^\prime $ are regarded equivalent if they differ only by a reparametrization, i.e., there is a linear map $h:V\rightarrow V$ such that
$$h(m^\prime(x,y))=m(h(x),h(y))\,,$$
for all $x,y$. Expanding to first order $h(x)=x+\delta(x)$ this means that $m^\prime$ and $m$ differ by a coboundary
\[m^\prime(x,y)-m(x,y)=m(\delta(x),y)+m(x,\delta(y))-\delta(m(x,y)),\]
specializing to $m_0$ this becomes
\begin{align}
    m(x,y)-x\cdot y=x\cdot \delta(y)-\delta(x\cdot y)+\delta(x)\cdot y\,. \label{cobdy}
\end{align}
The space of infinitesimal deformations of the product $m_0(-,-)$, or equivalently that of the associative algebra $A$, is given by the space of maps $\epsilon$ obeying \eqref{cocyc} modulo the image of coboundaries of the form \eqref{cobdy}. This is presicely the second Hochschild cohomology of $A$ as an $A$-bimodule, where $A$ acts on itself both from the left and from the right, usually denoted by $HH^2(A,A)$.

Note that the space of maps $\epsilon$ and $\delta$ are respectively the space of homomorphisms from tensor product of $A$ to $A$, by linearity. More formally and algorithmically, the Hochschild cohomology of $A$ is computed as the cohomology of the reduced bar complex 
\[0\rightarrow A\rightarrow \hom_{Vect_\C}(A,A)\rightarrow \hom_{Vect_\C}(A\otimes A,A)\rightarrow \hom_{Vect_\C}(A^{\otimes 3},A)\rightarrow \cdots\,,\]

where each term denotes the space of $\C$-multilinear maps from tensor product of $A$ to itself, respectively. The differential 
\begin{equation*}
d:\hom(A^{\otimes n},A)\rightarrow \hom(A^{\otimes n+1},A),
\end{equation*}
is given by \begin{align*}
(d\omega_n)(x_0,x_1,\cdots ,x_n)&=x_0\cdot \omega_n(x_1,\cdots ,x_n)\\
&+\sum_{i=1}^n(-1)^{i}\omega_n(x_1,\cdots ,x_{i-1}\cdot x_i,x_{i+1},\cdots ,x_n)+(-1)^{n+1}\omega_n(x_0,\cdots ,x_{n-1})\cdot x_n\,.
\end{align*}
\eqref{cobdy} and \eqref{cocyc} are concrete examples of the differential with $n=1,2$ respectively. In fact the differential is the alternating sum of the most general way of producing an element via a map that requires one fewer argument, preserving the order of inputs.

The notion of Hochschild cohomology can be generalized to more complicated algebraic structures, and to appropriate categories roughly by regarding the morphism spaces as an algebra. The category we will consider is the category of \textit{equivariant matrix factorizations} of a 2d $\mathcal N=(2,2)$ Landau-Ginzburg model. A pedagogical explanation of this is difficult; it suffices to know that it is the appropriate mathematical framework to describe the category of B-branes in this Landau-Ginzburg models with a weighted homogeneous superpotential.

\subsection{Berglund-H\"ubsch-Krawitz mirror symmetry for Landau-Ginzburg models}
Homological mirror symmetry (HMS) provides a powerful bridge between mirror Calabi-Yau geometries by identifying their categories of D-branes—namely, the derived Fukaya category of the A-model on one side and the derived category of coherent sheaves of the B-model on the mirror. For brevity, we will simply refer to them as the A-branes and B-branes.  

When the varieties under consideration develop singularities, the classical formulation of HMS does not apply directly. Defining the appropriate categories in the presence of geometric singularities is highly nontrivial. For instance, B-branes associated with singularities of algebraic varieties have been studied in the framework of \textit{singularity category} introduced by Orlov, see e.g. \cite{orlov2004triangulatedcategoriessingularitiesdbranes}, which roughly encode the local geometric data near the singularity. In some cases where the singularity is defined by an affine hypersurface, this category coincides with that of B-branes in the Landau–Ginzburg (LG) model defined by the same potential. On the A-side, there is by now a well-established version of HMS within the symplectic geometry community, involving the category of A-branes in the corresponding LG models. 

These developments suggest the following guiding idea: when studying the \textit{local geometry} of singularities, the relevant categories should be replaced by those of the underlying LG models defined by the same potential. This perspective is consistent with the general philosophy of string theory, where a geometric singularity is replaced by a nonsingular superconformal field theory - either a worldvolume CFT or an LG model. This viewpoint extends beyond the traditional Landau–Ginzburg/Calabi–Yau (LG/CY) correspondence \cite{Witten_1993,Vafa:1988uu} which concerns compact Calabi-Yau varieties\footnote{We thank the referee for carefully pointing this out.}. 

In the Landau–Ginzburg framework, B-branes are described by matrix factorizations, while A-branes correspond to vanishing cycles of a symplectic Lefschetz fibration. The only well-defined notion of mirror symmetry between LG models is that proposed by Berglund, Hübsch \cite{BERGLUND1993377} and later clarified by Krawitz \cite{krawitz2009fjrwringslandauginzburgmirror}. Originally, it was a combination of geometric mirror symmetry and LG/CY correspondence. Namely, starting from an LG model, one finds the corresponding compact Calabi-Yau 3-fold as the zero loci of superpotential in \textit{weighted projective spaces}. In the same way one may find the Calabi-Yau 3-fold corresponding to the Berglund-H\"ubsch mirror LG model. The claim is that these two Calabi-Yau 3-folds are mirror to each other in a suitable sense. This engineers many new mirror pairs and generalizes the earlier work of Greene and Plesser \cite{Greene:1990ud}

Although the pioneering construction of Berglund and H\"ubsch allows one to perform the mirror operation starting from the LG model, its interpretation is primarily geometric through LG/CY correspondence. The work of Krawitz, however, provides a fully intrinsic definition of mirror symmetry at the level of LG models themselves. We will not explain this, but turn to the basic construction below. Altogether, this represents a highly algebraic incarnation of mirror symmetry. 

Consider an \textit{invertible} polynomial $W$ of $n+1$ variables, which means a weighted homogeneous polynomial consisting of $n+1$ monomials
$$ W(x_1,x_2,\cdots ,x_{n+1})=\sum_{i=1}^{n+1}\prod_{j=1}^{n+1}x^{A_{ij}}_j\,,$$
where $A$ is a rank $n$ integral matrix with nonvanishing determinant. Taking the transpose of $A$ then yields the \emph{Berglund-H\"ubsch mirror} polynomial 
$$ \check{W}(x_1,x_2\cdots ,x_{n+1})=\sum_{i=1}^{n+1}\prod_{j=1}^{n+1}x^{A^T_{ij}}_j\,.$$
By construction, both $W$ and $\check W$ admit a natural 
$\mathbb{C}^*$–action. One fixes a system of positive integer weights $(d_1,\cdots,d_{n+1};h)$ with $g.c.d.(d_1,d_2,\cdots ,d_{n+1},h)=1$, so that
$$ W(\lambda^{d_1}x_1,\cdots ,\lambda^{d_{n+1}}x_{n+1})=\lambda^h W(x_1,\cdots ,x_{n+1})\,\,(\forall\lambda\in \C^\ast).$$
The requirement that the corresponding singularity remains at finite distance in the moduli space \cite{Gukov_2000} $\hat c=\sum_{i=1}^4(1-2q_i)<2$ is equivalent to the following condition
$$ d_0:=h-\sum_{i=1}^{n+1}d_i<0\,.$$
Beyond this 
$\mathbb{C}^*$ –action, W often enjoys a larger, finite symmetry group. One convenient description introduces an auxiliary coordinate 
$t_0$ and defines  
$$ \Gamma_W:=\left\{(t_0,t_1,\cdots ,t_{n+1})\in (\C^\ast)^{n+2}: \prod_{j=1}^{n+1}t_j^{A_{ij}}=t_0t_1t_2\cdots t_{n+1}\right\}\,.$$
This group acts on both $\C^{n+1}$ via multiplication by $(t_1,\cdots ,t_{n+1})$ and $\C^{n+2}$ via mulitplication by $(t_0,\cdots ,t_{n+1})$. The former action clearly preserves $W$, in the sense that the outcome is a $W$ up to a multiplicative factor. It preserves $W+x_0x_1\cdots x_{n+1}$ as well, regarded as a polynomial in $\C^{n+2}$ coordinatized by $(x_0,x_1,\cdots ,x_{n+1})$.

The homological mirror symmetry conjectures for Berglund-H\"ubsch mirror pairs (which is a special case of Berglund-H\"ubsch-Krawitz mirror) can be stated as follows (See e.g. \cite{evans2022symplecticcohomologycompounddu})

\begin{tcolorbox}
\begin{conj}
    There is a quasi-equivalence of idempotent complete $A_{\infty}$-categories
\begin{align}
    \mathcal F(\check{W})\simeq \text{mf}(\C^{n+1},\Gamma_W,W)
\end{align} 
between the Fukaya-Seidel category of a Morsification of $\check{W}$ and the dg-category of $\Gamma_W$-equivariant matrix factorizations of $W$.
\end{conj}

\begin{conj}
    There is a quasi-equivalence of idempotent complete $A_{\infty}$-categories
\begin{align}
    \mathcal W(\check{W}^{-1}(1))\simeq \text{mf}(\C^{n+2},\Gamma_W,W+x_0x_1\cdots x_{n+1})
\end{align} 
between the wrapped Fukaya category of the Milnor fiber $\check{W}^{-1}(1)$ and the dg-category of $\Gamma_W$-equivariant matrix factorizations of $W+x_0x_1\cdots x_{n+1}$.
\end{conj}
\end{tcolorbox}

Modulo technicality, this is a homological mirror symmetry statement establishing the equivalence of A-branes in the LG model $(\C^{n+1},W)$  and B-branes in the LG model $(\C^{n+1},\Gamma_{\check W},\check W)$. It turns out that for computational convenience it is recommended to replace the latter by $(\C^{n+2},\Gamma_{\check W},\check W+x_0x_1\cdots x_{n+1})$, introducing the extra $x_0$-coordinate. Using either of these conjectures and the fact \begin{align}
    SH^\ast(M_X)\cong HH^\ast(\mathcal W(M_X)), \label{shtohh}
\end{align}
it was proved in \cite{evans2022symplecticcohomologycompounddu} that if $$HH^2(\text{mf}(\C^{n+2},\Gamma_W,W))=0,$$ then there is an isomorphism
\begin{align}
    SH^\ast(\check{W}^{-1}(1))\cong HH^\ast(\text{mf}(\C^{n+2},\Gamma_W,W)). \label{shhh}
\end{align} 
Note that \eqref{shtohh} establish the symplectic-geometric invariant on the left to the more algebraic object on the right. 
 
Therefore, once we can prove that $HH^2(\text{mf}(\C^{n+2},\Gamma_W,W))=0$ for the singularity $W$, we can compute all their relevant symplectic cohomology groups via Hochschild cohomology at the mirror side. 
The latter is computable, albeit often very tedious. 
In particular, for isolated Gorenstein cDV singularities defined by invertible polynomials, one already knows that
$$
SH^k(M_X)=0,\text{ for }k=2\text{ or } k\geq 4.
$$
So the isomorphism \eqref{shhh} is fully consistent with these vanishing results.

We will make use of the formula (\ref{formula}) to determine the Hochschild cohomology of the equivariant matrix factorizations \cite{lekili2021homologicalmirrorsymmetrymilnor}. From a physics perspective its role is computing the Hilbert spaces of a Landau Ginzburg orbifold. Before we state the formula, let us introduce the necessary notations. 

As we have mentioned, elements in $\Gamma_W$ act canonically on the affine coordinates via coordinate-wise multiplications 
$$(t_0,\cdots ,t_{n+1})\cdot(x_0,x_1,x_2,\cdots ,x_{n+1})= (t_0x_0,t_1x_1,\cdots ,t_{n+1}x_{n+1})\,.$$
The character $\chi$ of the group $\Gamma_W$ is defined as
\begin{align}
    \Gamma_W\rightarrow \C^\ast, \qquad \chi (t_0,t_1,\cdots ,t_{n+1})=t_0t_1\cdots t_{n+1}=\prod_{j=1}^{n+1}t_j^{A_{ij}}\,.  \label{char}
\end{align} 
The character records an overall factor of each monomial contained in $W$ when acted on by $(t_0,\cdots ,t_{n+1})\in \Gamma_W$. The kernel of $\chi$ is the finite subgroup
$$ \text{ker}\chi=\left\{(t_0,\cdots ,t_{n+1})\in (\C^\ast)^{n+2}: \prod_{j=1}^{n+1}t_j^{A_{ij}}=1, t_0=t_1^{-1}\cdots t_{n+1}^{-1}\right\}\,.$$
Under this action, each element $\gamma\in\text{ker}\chi$ splits the coordinate space $V=\mathbb{C}^{n+2}$ (with basis $x_0,\cdots,x_{n+1}$) into its fixed subspace $V_\gamma$ and its complement subspace $N_\gamma$ in $V$. Given a choice of $\gamma\in $ker$\chi$, restricting $W$ to the set $V_\gamma$, i.e., the set of $\gamma$-fixed variables, is denoted $W_\gamma$. Let $\text{Jac}_{W_\gamma}$ be the associated Jacobian ring of $W_\gamma$. One picks a basis of the Jacobian ring as a vector space and label it by $J_\gamma$, this choice is eventually immaterial on the final constructions. 

Introduce dual coordinates $x_i^{\vee}, \text{for }i=0,1,\cdots ,n+1$, on which $\Gamma_W$ acts by
$$(t_0,\cdots ,t_{n+1})\cdot x_i^{\vee}=t_i^{-1}x_i^\vee\,. $$
Equivalently, the character of $x_i^{\vee}$ is inverse the character of $x_i$. For a general monomial $\underline{m}:=\prod_i x_i^{b_i},\text{ where } b_i\geq -1$ and each factor with $b_i=-1$ represents $x_i^{\vee}$, one can compute its character $\chi_{\underline{m}}$ by multiplying together all factors :
$\chi_{\underline{m}}=\prod_i t_i^{b_i}$.

Now we can state the formula, which reads \cite{lekili2021homologicalmirrorsymmetrymilnor}
\begin{equation}\label{formula}
\begin{aligned}
    HH^t(\text{mf}(\C^{n+2},\Gamma_W,W))\cong &\bigoplus_{\substack{\gamma\in \text{ker}\chi, l\geq 0, \\ t-\text{dim}N_\gamma=2u}}\left(H^{-2l}(d W_\gamma)\otimes \Lambda^{\text{dim}N_\gamma}N_\gamma^\vee\right)_{(u+l)\chi}\oplus \\ &\bigoplus_{\substack{\gamma\in \text{ker}\chi, l\geq 0, \\ t-\text{dim}N_\gamma=2u+1}}(H^{-2l-1}(dW_\gamma)\otimes \Lambda^{\text{dim}N_\gamma}N_\gamma^\vee)_{(u+l+1)\chi}\,,
\end{aligned}
\end{equation}
where $H^\ast(dW_\gamma)$ denotes the cohomology of the accosiated Koszul complex, $\Lambda^{\text{dim}N_\gamma}N_\gamma^\vee$ is the top exterior power of $N_{\gamma}^{\vee}$. Although the Koszul complex looks horrible, its cohomology actually concentrates at at most two degrees when the singularity $W_\g$ is isolated. More precisely, for each $\gamma\in $ ker $\chi$, there are two possible situations:

\begin{enumerate}
    \item $x_0$ is not fixed by $\gamma$: In this case, $W_\gamma$ has an isolated critical point at the origin. The cohomology of Koszul complex is concentrated in degree 0 which is isomorphic to $\text{Jac}_{W_\gamma}$, hence only the term $l=0$ in the first direct sum contributes to the summand. Choosing a basis $J_{\gamma}$ of Jacobian rings $\text{Jac}_{W_\gamma}$, each contributing element can be written as $$\underline{m}=p x_0^{\vee}x_{j_1}^{\vee}\cdots x_{j_{n+1-k}}^{\vee},$$ where $k$ is the number of fixed coordinates among $\{x_1,\cdots,x_{n+1}\}$, $x_{j_1},\cdots ,x_{j_{n+1-k}}$ label the unfixed coordinates and $p\in J_\gamma$ \footnote{Note that we do not care about the overall sign since eventually only the rank of $HH^t$ counts.}. One then imposes the character condition $\chi_{\underline{m}}=u\chi$ and sums over all integers $u$. 
    \item $x_0$ is fixed by $\gamma$:
   Only the terms $l=0$ contribute to the summand, but one for each direct sum. Monomials contributing to the first summand ($t-\text{dim}N_\gamma=2u$) are schematically $\underline{m}=x_0^{b_0}px_{j_1}^{\vee}\cdots x_{j_{n+1-k}}^{\vee}$ such that $\chi_{\underline{m}}=u\chi$. Monomials contributing to the second summand ($t-\text{dim}N_\gamma=2u+1$) are $\underline{m}=x_0^{b_0}px_0^{\vee}x_{j_1}^{\vee}\cdots x_{j_{n+1-k}}^{\vee}$ such that $\chi_{\underline{m}}=u\chi$. Here $b_0$ is a non-negative integer and $p\in J_\gamma$.
\end{enumerate}
Finally, one sum over all the elements $\gamma\in $ker$\chi$ and all allowed integers $u$ be derive $HH^t(\text{mf}(\C^{n+2},\Gamma_W,W))$.

In summary, there are three types of monomials that may potentially contribute to the Hochschild cohomology for a given $\gamma\in $ ker$\chi$. These are called $A_\gamma, B_\gamma, C_\gamma$ in \cite{evans2022symplecticcohomologycompounddu}
\begin{align*}
    A_\gamma &= \begin{cases}
                 \{x_0^\beta p x_{j_1}^{\vee} \cdots x_{j_{n+1-k}}^{\vee}\ :
                 \ p\in J_\gamma,\ \beta=0,1,2,\ldots\} & \mbox{ if }x_0\mbox{ is fixed by }\gamma \\
                 \emptyset&\mbox{ otherwise}
               \end{cases},\\
    B_\gamma &= \begin{cases} \{x_0^\beta p x_0^{\vee} x_{j_1}^{\vee}
                  \cdots x_{j_{n+1-k}}^{\vee}\ :
                  \ p\in J_\gamma,\ \beta=0,1,2,\ldots\} & \mbox{ if }x_0\mbox{ is fixed by }\gamma \\
                  \emptyset&\mbox{ otherwise}
               \end{cases},\\
    C_\gamma &= \begin{cases}
                 \emptyset & \mbox{ if }x_0\mbox{ is fixed by }\gamma \\
                 \{p x_0^{\vee} x_{j_1}^{\vee}\cdots x_{j_{n+1-k}}^{\vee}\ :
                 \ p\in J_\gamma\}&\mbox{ otherwise}
               \end{cases}.
  \end{align*}
  These monomials $\underline{m}$ will be called "good" once their characters obey $\chi_{\underline{m}}=u\chi$. Each good $A_\gamma$ monomial contributes rank one to $HH^{2u+n-k+1}$; each good $B_\gamma$ and $C_\gamma$ monomial contributes rank one to $HH^{2u+n-k+2}$, where $k$ is the number of coordinates in $\{x_1,\cdots ,x_{n+1}\}$ that are fixed by $\gamma$.

  The problem of determining ranks of corresponding Hochschild cohomology groups is then reduced to the problem of counting solutions to a set of integral linear (congruence) equations. In the next section, we will apply this tool to compute the symplectic cohomology of the candidate singularities in Problem  \ref{problem}. As we are considering singularities in $\C^4$, $n$ is specialized to be 3.

\paragraph{The example of $cA_1$:} To illustrate how the machinery works, let us recall a simple $cA_1$ example discussed in section 8.4 of \cite{peters2024isolatedhypersurfacesingularitiesalgebrageometric}. This singularity is defined by the invertible polynomial $W=x_1^2+x_2^2+x_3^2+x_4^2$. On the physics side, it admits a superconformal quiver gauge theory dual - namely, the Klebanov-Witten quiver with its standard superpotential \cite{Klebanov:1998hh}. Mathematically, the singularity has a crepant resolution with a single irreducible exceptional curve. The Berglund-H\"ubsch mirror is given by the same potential $\check W=W$, orbifolded by
  \[\Gamma_{\check W}=\{(t_0,\cdots ,t_4)|t_1^2=t_2^2=t_3^2=t_4^2=t_0t_1\cdots t_4\}\subset (\C^\ast)^5\,.\]
The character $\chi$ sends an element to $t_0t_1\cdots t_4$. The group ker$\chi$ as above 
\[G=\Z_2\times \Z_2\times \Z_2\times \Z_2\,,\]
where each $\Z_2$ factor contains two choices of $t_i$, $i=1,2,3,4$. $t_0$ can be eliminated. This gives a total of 16 elements in $G$. Let us see what subspaces they fix in $\C^{5}$. For example, $(-1,1,1,1)$ sends $(x_0,x_1,x_2,x_3,x_4)$ to $(-x_0,-x_1,x_2,x_3,x_4)$, hence it fixes $\C^3$ spanned by the last three coordinate lines, and $N_\g=$span$\{x_0,x_1\}$. We have
\[\check W_\g=x_2^2+x_3^2+x_4^2,\]
so $\text{Jac}_{W_\gamma}\cong\C$. One can take $J_\g=\{1\}$. As $x_0$ is not fixed by $\g$, the above rules mean that there is only one element $x_0^\vee x_1^\vee$ with character $t_0^{-1}t_1^{-1}$. This is a $C_\g$ monomial but not a good one, since its character is not in general a multiple of $\chi$. We conclude that this element $\g$ does not contribute to the Hochschild cohomology. Similar analysis revealed that there is no good $C_\g$ monomial. In fact the good condition forces such a monomial to be $x_0^\vee x_1^\vee x_2^\vee x_3^\vee x_4^\vee$, which can never appear as a $C_\g$ monomial. The only possible choice of $\g$ that fixes none of $x_1,\cdots,x_4$ is $(-1,-1,-1,-1)$, which fixes $x_0$. However this is actually a good $B_\g$ monomial, and it contributes to $HH^3(\C^{5},\G_{\check W},\check W)$. This is the only contribution to $HH^3$. This agrees with the fact that $SH^3$ of the Milnor fiber of the singularity defined by $W$ is of rank 1, the Milnor number of the singularity.

Now we turn to $A_\g$ and $B_\g$ monomials. This requires considering $\g$ that fixes $x_0$. Observe that $\text{Jac}_\g\cong \C$ in all situations, where we remove $x_0$ in the space $V_\g$ where $W_\g$ is defined. Hence we take $J_\g=\{1\}$. One finds that the only good $A_\g$ monomials are the following:
\begin{itemize}
    \item  $x_0^\b x_1^\vee x_2^\vee x_3^\vee x_4^\vee$ for $\g=(-1,-1,-1,-1)$, with $\b=1,3,5,\cdots$. In fact when $\b=2k-1$ this monomial can be rearranged to be $x_0^{-2k-1}x_0^{4k}x_1^\vee x_2^\vee x_3^\vee x_4^\vee$, which has character $\chi_{\underline{m}}=\chi^{\otimes (-2k-1)}$, or simply denoted as $\chi_{\underline{m}}=(-2k-1)\chi$. Hence it contributes rank one to the Hochschild cohomology at degree $2-4k$.
    \item $x_0^\b$ for $\g=(1,1,1,1)$, with $\b=0,2,\cdots$. When $\b=2k$ this has character $\chi_{\underline{m}}=-2k\chi$. Hence it contributes rank one to the Hochschild cohomology at degree $-4k$.
\end{itemize}
Altogether these contribute to the Hochschild cohomology group once at each non-positive even degree. Now the good $B_\g$ monomials other than the one discussed above can be obtained in a similar way, i.e.,
\begin{itemize}
    \item  $x_0^\b x_0^\vee x_1^\vee x_2^\vee x_3^\vee x_4^\vee$ for $\g=(-1,-1,-1,-1)$, with $\b$ a non-negative even number. When $b=2k$. This has character $\chi_{\underline{m}}=(-2k-1)\chi$. Hence it contributes rank one to the Hochschild cohomology at degree $3-4k$.
    \item $x_0^\b x_0^\vee$ for $\g=(1,1,1,1)$, with $\b$ a positive odd number. When $\b=2k+1$ this has character $\chi_{\underline{m}}=-2k\chi$. Hence it contributes rank one to the Hochschild cohomology at degree $1-4k$.
\end{itemize}
Altogether these contribute to the Hochschild cohomology group once at all odd degree not greater than 1. As a result we obtain 
\begin{align}
    \begin{cases}
        HH^3(\text{mf}(\C^5,\Gamma, \check{W}))=1\,, \\
        HH^{d\leq 1}(\text{mf}(\C^5,\Gamma, \check{W}))=1\,, \\
        HH^{d=2\text{ or }d\geq 4}(\text{mf}(\C^5,\Gamma, \check{W}))=0\,,
    \end{cases}
\end{align}
which is in agreement with \eqref{shpositive}. It is important to bear in mind that a given monomial may appear as either a $A_\g,B_\g$ or $C_\g$ monomial.

\section{Computation of symplectic cohomology of $cE_6$ and $cE_7$ singularities} \label{hos}
Our goal in this section is to compute symplectic cohomology of (Milnor fiber of) the singularities $W=0$ in Problem \ref{problem}. Once the defining polynomials are all invertible, one can apply the techniques from the mirror side developed above. As a first step, one must verify that $HH^2(\text{mf}(\C^5,\G_{\check{W}},\check{W}))$ vanish. 

\subsection{$cE_6$ singularities}
These singularities are
$$W=x^2+y^3+z^4+y w^k:  1<k<16\,.$$
The Berlund-H\"ubsch mirrors are (after relabeling the coordinates)
$$\check{W}=x^2+y^4+z^3w+w^k:  1<k<16\,.$$
Our goal here is to show that for these mirror singularities the Hochschild cohomology satisfies
$$
HH^2(\text{mf}(\C^5,\G_{\check{W}},\check{W}))=0
$$ and moreover $HH^t$ stablizes for $t<0$ if and only if $k=8$\,.
\paragraph{Example $k=8$:}
This case is a bit more difficult than the computation of Brieskorn-Pham singularities considered in \cite{evans2022symplecticcohomologycompounddu}, due to the mixing of $\C^\ast$-actions on $z$ and $w$. We follow and slightly generalize a method proposed in \cite{adaloglou2024symplecticcohomologyquasihomogeneouscan}. We illustrate the method for $k=8$ as follows.

First, one can compute the normalized weights: $(d_0,\cdots,d_4)=(-4,12,6,7,3)$ and $h=24$. The basis of Jacobian algebra is taken to be 
\begin{align}
J=
\begin{rcases}
\begin{dcases}
    &1,z,z^2,w,\cdots,w^7,zw,\cdots,zw^7 \quad \text{ if both }z, w\text{ are fixed.} \\
    &1,w,\cdots,w^6\quad \text{ if }w\text{ is fixed but }z\text{ is not.} \\
    & 1 \quad \text{ if }w\text{ is not fixed.}
\end{dcases}
\end{rcases}
    \bigotimes
    \begin{rcases}
\begin{dcases}
    &1 \quad \text{ if }y\text{ is not fixed.} \\
    &1,y,y^2\quad \text{ if }y\text{ is fixed.} 
\end{dcases}
\end{rcases}\,.
\end{align}

In the presence of the mixing of $\C^\ast$-actions, one can simplify the problem by lifting the group $\Gamma_{\check{W}}$ to its covering. To simplify notation, we write $\Gamma$ for  $\Gamma_{\check{W}}$ in the following. We define a map $\Psi$ 
\begin{align}
    \Psi:(\C^\ast)^4\rightarrow \Gamma, \quad (u_1,u_2,u_3,\t)\mapsto (\tau^{d_0}u_1^{-1}u_2^{-1}u_3^2,\t^{d_1}u_1,\t^{d_2}u_2,\t^{d_3}u_3,\t^{d_4}u_3^{-3})=(t_0,t_1,\cdots,t_4)\,.\label{hto1}
\end{align}
This map is certainly surjective. Demanding that the image is contained in $\G$ imposes the following 
$$ u_1^2=u_2^4=u_3^{24}=1\,.$$
With this new parametrization, the character $\chi\circ \P=\t^{h}$. The kernel of map $\P$ is precisely $\Z_h$.
Now consider the ker($\chi \circ\P)=\Z_2\times \Z_4\times \Z_{24}\times \Z_h$, hence ker$\chi=\Z_2\times \Z_4\times \Z_{24}$ with generators the roots of unity. An element $\g\in$ ker$\chi$ acts on coordinates via $(x_0,\cdots,x_4)\mapsto(u_1u_2^3u_3^2x_0,u_1x,u_2y,u_3z,u_3^{-3}w)$ where we have made an identification $x_1=x,x_2=y,x_3=z,x_4=w$.

Now one can compute the character of any monomial $\underline{m}=x_0^{b_0}x^{b_1}\cdots w^{b_4}$ to be $\chi_{\underline{m}}=\t^{n_0}u_1^{n_1}u_2^{n_2}u_3^{n_3}$ where 
\begin{align}
    \begin{cases}
        n_0&=-4b_0+12b_1+6b_2+7b_3+3b_4\,,\\
        n_1&=b_1-b_0\,,\\
        n_2&=b_2-b_0\,,\\
        n_3&=-3b_4+b_3+2b_0\,.
    \end{cases} \label{constr1} 
\end{align}
The existence of a corresponding $\g\in$ ker$\chi$ then means: $b_i=0$ if $x_i$ is fixed by $\g$; $\prod_{j\in I\subset\{1,2,3,4\}} x_j\in J_\g$ if $x_{j\in I}$ are coordinates fixed by $\g$. Recall the order of $u_i$, we see that the condition for $\underline{m}$ to be "good" is 
\begin{align}
    2|n_1,\quad 4|n_2,\quad 24|n_3,\quad 24|n_0\,. \label{constr2} 
\end{align}
Then one concludes that $u=n_0/h$. This monomial will hence contribute to the Hochschild cohomology at a certain degree, depending on $u$ and on which type it belongs to.

In principle, the algorithm for computing the rank at a certain value of $u$ is the following: One has to enumerate all $\g$, first identifying whether it fixes $x_0$ or not to cast the corresponding monomials into one of the three types. Then one enumerates on all monomials in the Jacobian algebras of the fixed variables, and apply constraints \eqref{constr1},\eqref{constr2} to find number of integral solutions $(b_0,b_1,\cdots,b_4)$.

In the case at hand, there is a short-cut. Observe that $b_1\equiv b_2$ (mod2). The only possibilities for $(b_1,b_2)$ are thus $(-1,-1),(-1,1),(0,0)$ and $(0,2)$. First we discuss the $A$-type monomials. There are only few choices of $\g$. Let $\z$ be an $24-$th root of unity.
\begin{enumerate}
    \item $\g=(1,1,1)$: In this case, $z,w$ are fixed, so they appear as monomials in the Jacobian. \begin{itemize}
        \item If $b_1=b_2=0$, the constraints imply that $4|b_0$ and $24|(b_0+b_3-3b_4)$. A computation then shows that $b_0\equiv 0,4$ (mod 12). There are two possible monomials, $x_0^{12k}$ and $x_0^{12k+4}zw^3$, which contribute two generators in $HH^{-4k}$ ($k\geq 0$);
        \item    If $b_1=0,b_2=2$, the constraints imply that $4|(b_0-2)$ and $24|(b_0+b_3-3b_4)$. One concludes that $b_0\equiv 6,10$ (mod 12). The monomials $x_0^{12k+6}y^2w^4$ and $x_0^{12k+10}y^2zw^7$ contribute two generators in $HH^{-4k}$.
    \end{itemize}
    
    \item $\g=(1,1,-1)$: In this case, $z,w$ are not fixed, so they appear as $z^\vee w^\vee$. \begin{itemize}
        \item If $b_1=b_2=0$, the constraints imply that $4|b_0$ and $24|(b_0+2)$. There is no solution;
        \item    If $b_1=0,b_2=2$, the constraints imply that $4|(b_0-2)$ and $24|(b_0+2)$. Again, no solution.
    \end{itemize}
    \item $\g=(-1,1,\pm i)$: $b_1=-1$, $b_2=1$. The constraints impose that $4|(b_0-1)$ and $24|(2b_0+2)$. There is no solution.
    \item $\g=(-1,-1,1)$: $b_1=b_2=-1$, $4|(b_0+1)$, $24|(2b_0+b_3-3b_4)$. One concludes that $b_0\equiv -1,3$ or 7 (mod 12). The case $b_0\equiv -1$ (mod 12) requires extra care. When $b_0=-1$ there is a good monomial $x_0^\vee x^\vee y^\vee z^2$. It contributes to $HH^3$. Otherwise, there are three monomials $x_0^{3+12k}x^\vee y^\vee w^2,x_0^{7+12k}x^\vee y^\vee zw^5, x_0^{11+12k}x^\vee y^\vee z^2$ where $k\geq 0$. Two of them contribute to $HH^{-4k}$ and the remaining one contributes to $HH^{-4k-2}$.
    \item $\g=-1,-1,-1$: $b_1=b_2=-1$, $4|(b_0+1)$, $24|(2b_0+2)$. One concludes that $b_0\equiv -1$ (mod 12). The case $b_0=-1$ yields a good monomial $x_0^\vee x^\vee y^\vee z^\vee w^\vee$. It contributes to $HH^3$. Otherwise, $x_0^{11+12k}x^\vee y^\vee z^\vee w^\vee$ contributes to $HH^{-4k-2}$.
    \item $\g=(-1,i,\pm \z^9)$: $4|(b_0+1)$, $24|(2+2b_0)$. One concludes that $b_0\equiv -1$ (mod 12). For the same reason, both $HH^3$ and $HH^{-4k-2}$ acquire two generators, since there are two possible choices of $\g$.
    \item $\g=(-1,-i,\pm \z^3)$: The same as in case 6.
\end{enumerate}

To summarize, one obtains Hochschild cohomology groups of rank 6 at all nonpositive even degrees, and also at degree 3. The contributions to $HH^3$ are in fact due to $B$-type monomials. Nevertheless, one performs the following trick: Write $1=x_0 x_0^\vee$ and insert it to the $A$-type monomials obtained above. This produces all the remaining $B$-type monomials. One finds that they contribute exactly to Hochschild cohomology groups at one degree higher than the corresponding $A$-monomials. In this way, one obtains $HH^t$ of rank 6 at all odd degree less than or equal to one.

It remains to analyze the $C$-type monomials. There are 66 possible choices of $\g$ that does not fix $x_0$; each of them may lead to one $C$-type monomial. 2 of them fix $z,w$. 6 of them fix $w$ without fixing $z$, while the others fix neither of $z$ and $w$. One finds that, except for these 6 elements, all 66 elements contribute one generator respectively to $HH^3$. In total, we obtain
\begin{align}
    \begin{cases}
        HH^3(\text{mf}(\C^5,\Gamma, \check{W}))=66\,, \\
        HH^{d\leq 1}(\text{mf}(\C^5,\Gamma, \check{W}))=6\,, \\
        HH^{d=2\text{ or }d\geq 4}(\text{mf}(\C^5,\Gamma, \check{W}))=0\,.
    \end{cases}
\end{align}
At positive degrees, this coincide with the known symplectic cohomology of the original singularities $W=0$. The second Hochschild cohomology of $\check{W}=0$ vanishes as expected, and so \eqref{shhh} can be applied. As a sanity check, we see that the Milnor number of $W=0$ is precisely 66. We then conclude that the symplectic cohomology groups of the Milnor fiber of $W^{-1}(0)$ have rank 6 at all negative degrees.

\paragraph{General values of $k$:}
With the experience of dealing with $k=8$, it is now straightforward to compute the Hochschild cohomology for general values of $k$. 

We do not have to evaluate $d_i$ in general, since only the proportion $\w_i=\frac{d_i}{h}$ matters. We have $(\w_0,\cdots,\w_4)=(\frac{1}{4}-\frac{k+2}{3k},\frac{1}{2},\frac{1}{4},\frac{k-1}{3k},\frac{1}{k})$. The $h:1$ covering homomorphism is given by
\eqref{hto1} again, but the kernal of $\chi$ is now
$$ u_1^2=u_2^4=u_3^{3k}=1\,.$$
It acts via $(x_0,\cdots,x_4)\mapsto(u_1u_2^3u_3^2x_0,u_1x,u_2y,u_3z,u_3^{-3}w)$.
The choice of monomial basis is the following
\begin{align}
J=
\begin{rcases}
\begin{dcases}
    &1,z,z^2,w,\cdots,w^{k-1},zw,\cdots,zw^{k-1} \quad \text{ if both }z, w\text{ are fixed.} \\
    &1,w,\cdots ,w^{k-2}\quad \text{ if }w\text{ is fixed but }z\text{ is not.} \\
    & 1 \quad \text{ if }w\text{ is not fixed.}
\end{dcases}
\end{rcases}
    \bigotimes
    \begin{rcases}
\begin{dcases}
    &1 \quad \text{ if }y\text{ is not fixed.} \\
    &1,y,y^2\quad \text{ if }y\text{ is fixed.} 
\end{dcases}
\end{rcases}\,.
\end{align}

Good monomials obey the following conditions 
$$2|(b_1-b_0),\quad 4|(b_2-b_0),\quad 3k|(b_3-3b_4+2b_0)\,.$$
And it is required that $u=(\frac{1}{4}-\frac{k+2}{3k})b_0+\frac{b_1}{2}+\frac{b_2}{4}+\frac{(k-1)b_3}{3k}+\frac{b_4}{k}$ is integral. Again, there are four choices of the tuple $(b_1,b_2)$. The elements $\g$ that fix $x_0$ can be case into one of the five classes
\begin{enumerate}
    \item $\g=(1,1,1)$: In this case, $z,w$ are fixed, so they appear as monomials in the Jacobian. \begin{itemize}
        \item $b_1=b_2=0$. In all, number of generators contributed to $HH^{2u}$ is the number of integral tuples $(b_0,b_3,b_4)$ obeying the following constraints
        \begin{align*}
        \begin{cases}
            u=\left(\frac{1}{4}-\frac{k+2}{3k}\right)b_0+\frac{(k-1)}{3k}b_3+\frac{b_4}{k}\,, \\
            b_0\geq 0, \quad 4|b_0, \quad 3k|(2b_0+b_3-3b_4)\,, \\
            z^{b_3}w^{b_4}\in J\,.
        \end{cases}
        \end{align*}
        
        \item $b_1=0,b_2=2$. In all, number of generators contributed to $HH^{2u}$ is the number of integral tuples $(b_0,b_3,b_4)$ obeying the following constraints
        \begin{align*}
        \begin{cases}
            u=\left(\frac{1}{4}-\frac{k+2}{3k}\right)b_0+\frac{1}{2}+\frac{(k-1)b_3}{3k}+\frac{b_4}{k}\,, \\
            b_0\geq 0, \quad 4|(b_0-2), \quad 3k|(2b_0+b_3-3b_4)\,, \\
            z^{b_3}w^{b_4}\in J\,.
        \end{cases}
        \end{align*}
    \end{itemize}
    
    \item $\g=(1,1,-1)$: In this case, $z,w$ are not fixed, so they appear as $z^\vee w^\vee$. \begin{itemize}
        \item Number of generators contributed to $HH^{2u+2}$ is the number of integral tuples $(b_0,b_3,b_4)$ obeying the following constraints
        \begin{align*}
        \begin{cases}
            u=\left(\frac{1}{4}-\frac{k+2}{3k}\right)b_0-\frac{(k+2)}{3k}\,, \\
            b_0\geq 0, \quad 4|b_0, \quad 3k|(2b_0+2)\,, \\
            z^{b_3}w^{b_4}\in J\,.
        \end{cases}
        \end{align*}
        \item  Number of generators contributed to $HH^{2u+2}$ is the number of integral tuples $(b_0,b_3,b_4)$ obeying the following constraints
        \begin{align*}
        \begin{cases}
            u=\left(\frac{1}{4}-\frac{k+2}{3k}\right)b_0-\frac{(k+2)}{3k}+\frac{1}{2}\,, \\
            b_0\geq 0, \quad 4|b_0, \quad 3k|(2b_0+2)\,, \\
            z^{b_3}w^{b_4}\in J\,.
        \end{cases}
        \end{align*}
    \end{itemize}
    \item $\g=(-1,-1,1)$: Number of generators contributed to $HH^{2u+2}$ is the number of integral tuples $(b_0,b_3,b_4)$ obeying the following constraints
        \begin{align*}
        \begin{cases}
            u=\left(\frac{1}{4}-\frac{k+2}{3k}\right)b_0-\frac{3}{4}+\frac{(k-1)b_3}{3k}+\frac{b_4}{k}\,, \\
            b_0\geq 0, \quad 4|(b_0+1), \quad 3k|(2b_0-b_3+3b_4)\,, \\
            z^{b_3}w^{b_4}\in J\,.
        \end{cases}
        \end{align*}
        Note that for $b_0=-1$ there is an extra generator of $HH^3$, at $u=0$.
    \item $\g=(-1,-1,-1)$: Contributions to $HH^{2u+4}$ are from integral tuples $(b_0,b_3,b_4)$ obeying the following constraints
        \begin{align*}
        \begin{cases}
            u=\left(\frac{1}{4}-\frac{k+2}{3k}\right)b_0-\frac{3}{4}-\frac{k+2}{3k}\,,\\
            b_0\geq 0, \quad 4|(b_0+1), \quad 3k|(2b_0+2)\,, \\
            z^{b_3}w^{b_4}\in J\,.
        \end{cases}
        \end{align*}
        Note that for $b_0=-1$ there is an extra generator of $HH^3$, at $u=-1$.
    \item $\g=(-1,1,\pm i)$: This is only possible when 
$4|k$. Contributions to $HH^{2u+3}$ are from integral tuples $(b_0,b_3,b_4)$ obeying the following constraints
        \begin{align*}
        \begin{cases}
            u=\left(\frac{1}{4}-\frac{k+2}{3k}\right)b_0-\frac{1}{4}-\frac{k+2}{3k}\,,\\
            b_0\geq 0, \quad 4|(b_0-1), \quad 3k|(2b_0+2)\,, \\
            z^{b_3}w^{b_4}\in J\,.
        \end{cases}
        \end{align*}
    \item $\g=(-1,i,\pm \sqrt{-i})$ or $(-1,-i,\pm \sqrt{i})$: Contributions to $HH^{2u+4}$ are from integral tuples $(b_0,b_3,b_4)$ obeying the following constraints
        \begin{align*}
        \begin{cases}
            u=\left(\frac{1}{4}-\frac{k+2}{3k}\right)b_0-\frac{1}{4}-\frac{k+2}{3k}\,,\\
            b_0\geq 0, \quad 4|(b_0+1), \quad 3k|(2b_0+2)\,, \\
            z^{b_3}w^{b_4}\in J\,.
        \end{cases}
        \end{align*}
    Note that there are in total four extra tuples contributing to $HH^3$ for this class.
\end{enumerate}
As in the example where $k=8$, $B$-monomials contribute the same amount to cohomology groups at one degree higher (but not to $HH^4$). $C$-monomials contribute in total $9k-12$ generators to $HH^3(\text{mf}(\C^5,\G,\check{W}))$. Hence one finds that the rank at degree 3 is $9k-6$, which reproduces the Milnor number of $W$ as expected.

We implement the above computations using \texttt{Mathematica}. The results at degree $-10\leq d\leq 2$ are listed in Table  \ref{tab:cE6hh}.
\begin{table}[!h]
    \centering
    \begin{tabular}{|c|c|}\hline
        $k$ & rank $HH^d, d=-10,-9,\cdots ,1,2$ \\ \hline
        2 & $3,3,2,2,3,3,2,2,2,2,3,3,0$\\ \hline
        3 & $2,2,2,2,1,1,3,3,3,3,2,2,0$\\ \hline
        4 & $2,2,3,5,5,5,2,2,3,5,5,5,0$\\ \hline
        5 & $3,3,1,1,4,4,3,3,2,2,4,4,0$\\ \hline
        6 & $5,5,3,3,3,3,3,3,3,3,5,5,0$\\ \hline
        7 & $2,2,4,4,3,3,5,5,1,1,5,5,0$\\ \hline
        8 & $6,6,6,6,6,6,6,6,6,6,6,6,0$\\ \hline
        9 & $2,2,5,5,3,3,5,5,1,1,6,6,0$\\ \hline
        10 & $5,5,3,3,3,3,5,5,3,3,6,6,0$\\ \hline
        11 & $6,6,3,3,5,5,4,4,2,2,6,6,0$\\ \hline
        12 & $6,6,2,2,6,6,3,5,5,5,6,6,0$\\ \hline
        13 &$5,5,3,3,6,6,2,2,4,4,6,6,0$\\ \hline
        14 & $3,3,3,3,6,6,3,3,5,5,6,6,0$\\ \hline
        15 & $3,3,5,5,6,6,1,1,5,5,6,6,0$ \\\hline
    \end{tabular}
    \caption{$HH^d(\text{mf}(\C^5,\G,\check{W}))$ for $-10\leq d\leq 2$.}
    \label{tab:cE6hh}
\end{table}
In particular, we find that $HH^2(\text{mf}(\C^5,\G,\check{W}))=0$ for all $k$, and rank$(HH^{d<0}(\text{mf}(\C^5,\G,\check{W})))$ stablizes only if $k=8$, which is the case shown in detail in the previous example. Hence we assert that only in that case there exists a crepant resolution.

\subsection{$cE_7$ singularities}
  The singularities are
  $$W=x^2+y^3+y z^3+z w^{k}: 2\leq k\leq 26, 2|k\, . $$
  The corresponding Berglund-H\"ubsch mirrors are
  $$\check{W}=x^2+y^3z+z^3w+w^{k}: 2\leq k\leq 26, 2|k\, .$$
  Our goal is to show that the rank of Hochschild cohomology groups of these mirror singularities is zero at degree 2, and stablize at all negative degrees if and only if $k=2$ or $k=14$.
  
Hochschild cohomologies of these singularities are much more difficult to compute than the $E_6$ cases, due to the mixing of characters. We follow and slightly generalize the trick in \cite{evans2022symplecticcohomologycompounddu}. Define the group $G\subset\Z_2\times \Z_3\times (\C^\ast)^2$ which contains elements $(s,\mu,\r,\t)$ with relation $\r^3=\mu^2\t^{4k+2}$. This can be mapped surjectively to $\Gamma$ via
\begin{align}
    \Psi:G\rightarrow \Gamma, \quad (s,\mu,\r,\t)\mapsto (s\r^{-1}\mu^2\t^{k-4},s\t^{3k},\r,\mu\t^{2k-2},\t^6)=(t_0,t_1,\cdots ,t_4)\,.\label{6to1}
\end{align}
Now $\chi\circ \P=\t^{6k}$. One can see that ker$\P$ is the following subgroup
$$ \left\{\left(s=1,\mu=\t^{2-2k},\r=1,\t\right)\left|\t^6=1\right.\right\}. $$
This is an order 6 subgroup, so one concludes that the map $\P$ is 6 to 1. 

One picks the monomial basis 
\begin{align*}
J_{\g \text{ fixes }y,z,w}=&
    \begin{cases}
         y^l z^i w^j,\quad &l=0,1; i=0,1,2; j=0,1,\cdots ,k-2\,,\\
        y^2 w^j,\quad &j=0,1,\cdots ,k-2\,,\\
        y^l z^i w^{k-1},\quad &l=0,1; i=0,1,2\,,\\
    \end{cases}\\
    J_{\g\text{ fixes }z,w}=&\begin{cases}
         z^i w^j,\quad &i=0,1; j=0,1,\cdots ,k-1\,,\\
         z^2\,,
    \end{cases}\\
    J_{\g\text{ fixes }z,y}=&\begin{cases}
         y^i z^j,\quad &i=0,1; j\geq 0\,,\\
         y^2\,,
    \end{cases}\\
    J_{\g\text{ fixes }w}=&\begin{cases}
         w^j,\quad &j=0,1,\cdots ,k-1\,.
    \end{cases}
\end{align*}
We note that if $\g$ fixes $z,y$ but not $w$, then the singularity $\check{W}_\g=0$ is not isolated, and the simplification below \eqref{formula} may break down. However, as we will see, it is expected that the caveat is immaterial in the current context.

As before, consider first the case that $\g$ fixes $x_0$. This means the following
\begin{align}
    \begin{cases}
        s\t^{k-4}=\r \mu\,,\\
        \r^3=\mu^2\t^{4k+2}\,,\\
        \t^{6k}=1\,.
    \end{cases}
\end{align}
Eliminating $\r$ yields $s\mu=\t^{k+14}$ and in particular $\t^{84}=1$. Hence the details depend on the g.c.d. of $6k$ and 84.
Since $g.c.d.(6k,84)=12$ for all $k\neq 14$, we find that there are two possible $\g$, given by $\t^6=1$ and $\t^6=-1$\footnote{Recall that $\P$ is 6 to 1, so each possibility yields only one element $\g$.}, respectively, that fix $x_0$ for each such $k$. One such $\g$ fixes all coordinates, while the other fixes $x_0$ only. When $k=14$, $\t$ can only be determined up to $\t^{84}=1$, hence there are fourteen elements $\g$. One of them fixes all, six of them fix $x_0,x$ and the remaining seven fix $x_0$ only. In that case, one finds the following 
\begin{align}
    (s,\mu,\r,\t)=(1,\z^{4m}\t^4,\z^{11m},\t|\t^6=\z^m, m=0,1,\cdots ,13 )\, ,
\end{align}
where one picks a 14-th root of unity and let it be $\z$.

Hence, denote a monomial $\underline{m}=x_0^{b_0}x^{b_1}\cdots w^{b_4}$ as before, we can compute its character and look at the following 
\begin{enumerate}
    \item $\g$ fixes all variables: $\chi_{\underline{m}}=\chi^{\otimes u}$ means that, $s,\mu$ and $\rho$ must be eliminated by only applying the defining relations $s^2=\mu^3=1$ and $\r^3=\mu^2\t^{4k+2}$. Hence there are contributions to $HH^{2u}$ from integral tuples $(b_0,b_2,b_3,b_4)$ obeying the following constraints
        \begin{align*}
        \begin{cases}
            b_0\geq 0, \quad 2|b_0, \quad 3|(b_2-b_0),\quad 3\left|\left(2b_0+b_3+2\frac{b_2-b_0}{3}\right)\right.\,, \\
            6ku=\left(4k+2\right)\frac{b_2-b_0}{3}+b_0(k-4)+b_3(2k-2)+6b_4\,,\\
            x_0^{b_0}y^{b_2}z^{b_3}w^{b_4}\in J\,.
        \end{cases}
        \end{align*}
        
        \item $\g$ fixes $x_0$ only: $b_1=b_2=b_3=b_4=-1$. The character condition says that there are contributions to $HH^{2u+4}$ from the integer $b_0$ obeying the following constraints
        \begin{align*}
        \begin{cases}
            b_0\geq 0, \quad 2|(b_0-1), \quad 3|(1+b_0),\quad 3\left|\left(2b_0+2-2\frac{b_0+1}{3}\right)\right.\,, \\
            6ku=-\left(4k+2\right)\frac{1+b_0}{3}+b_0(k-4)-5k-4\,.
        \end{cases}
        \end{align*}
        
        \item $\g$ fixes $x_0$ and $x$ only: This is possible only when $k=14$. $b_2=b_3=b_4=-1$; $b_1$ can only be 0 in the Jacobian algebra. There are contributions to $HH^{2u+3}$ from integer $b_0$ obeying the following constraints
        \begin{align*}
        \begin{cases}
            b_0\geq 0, \quad 2|b_0, \quad 3|(1+b_0),\quad 3\left|\left(2b_0+2-2\frac{b_0+1}{3}\right)\right.\,,  \\
            6ku=-\left(4k+2\right)\frac{1+b_0}{3}+b_0(k-4)-2k-4, \quad u\text{ is integral}\,.
        \end{cases}
        \end{align*}
\end{enumerate}
These exhaust all possible $A$-monomials. Some remarks are in order. First of all, one can still perform the substitution to obtain $B$-monomials which contribute to one degree higher, but as in the $E_6$ case there can be extra $B$-monomials with $b_0=-1$. To find them, note that as the character of $x_0$ contains $s$, $\chi_{\underline{m}}$ depends merely on $\t$ only if $b_0$ is even or $b_0$ is odd with $b_1=-1$. If one further demands that $\g$ fixes $x_0$, then the only source for those extra $B$-monomials are from class 2 of the above, with $\underline{m}=x_0^\vee x^\vee y^\vee z^\vee w^\vee$. Each such $\g$ contributes one generator to $HH^3$. Therefore we exhaust all possible $A$- and $B$-monomials from this procedure. 

A second remark is that, by implementing the above computations in \texttt{Mathematica}, we found that no $A$- and $B$-monomial obtained in the procedure above contributes to $HH^2(\text{mf}(\C^5, \G,\check{W}))$, as one may expect. On the other hand, stabilization at negative degrees happens precisely when $k=2$ and $k=14$. Class 3 in fact contributes nothing; This is desired since it yields generators of cohomology at odd degrees.

It remains to analyze $C$-monomials. We will show that they contribute only to $HH^3(\text{mf}(\C^5, \G,\check{W}))$. We do not aim to count the precise rank of $HH^3$; In fact, it suffices to ensure that they do not contribute to $HH^2$ in order for \eqref{shhh} to hold. 

Again, we work by enumeration. To obtain a good monomial, $\g$ must not fix $x$ once it does not fix $x_0$. Observe from \eqref{6to1} that once $\g$ fixes $y$ and $w$ it must then fix $z$. 
The constraints read 
\begin{align*}
        \begin{cases}
            b_0=-1, \quad 3|(b_2+1),\quad 3\left|\left(1+b_3+2\frac{b_2+1}{3}\right)\right.\,, \\
            6ku=\left(4k+2\right)\frac{b_2+1}{3}+4-4k+b_3(2k-2)+6b_4,\quad u\text{ is integral}\,.
        \end{cases}
        \end{align*}

There are the following cases
\begin{enumerate}
    \item $\g$ fixes $y$ only. One concludes that $\r=1$. The only possible $\underline{m}$ appears $x_0^\vee x^\vee z^\vee w^\vee$. $b_0=0$ violates the constraints.
    \item $\g$ fixes $y,z$. $\underline{m}=x_0^\vee x^\vee y^2 w^\vee$. This contributes to $HH^3$.
    \item $\g$ fixes $y,z,z$. $\underline{m}=x_0^\vee x^\vee y^2 w^{i}$ where $i=0,1,\cdots ,k-2$. Then $u=(b_4+1)/k$ cannot be integral.
    \item $\g$ fixes $z$ only. $\underline{m}=x_0^\vee x^\vee y^\vee w^\vee$. Constraints violated by $b_3=0$.
    \item $\g$ fixes $z,w$. It is only possible that $\underline{m}=x_0^\vee x^\vee y^\vee z^2$. This contributes to $HH^3$.
    \item $\g$ fixes $w$ only. $\underline{m}=x_0^\vee x^\vee y^\vee z^\vee w^i$ where $i=0,1,\cdots ,k-2$. No integral $u$ exists.
    \item $\g$ fixes nothing. $\underline{m}=x_0^\vee x^\vee y^\vee z^\vee w^\vee$. One finds that $u=1$; this contributes to $HH^3$.
\end{enumerate}

Combined with the remarks in the previous paragraph, we assert that all the $cE_7$ singularities in Problem \ref{problem} admit no crepant resolutions. The results at degree $-10\leq d\leq 2$ are listed in Table  \ref{tab:cE7hh}. 

\begin{table}[!h]
    \centering
    \begin{tabular}{|c|c|}\hline
        $k$ & rank $HH^d, d=-10,-9,\cdots ,1,2$ \\ \hline
        2 & $1,1,1,1,1,1,1,1,1,1,1,1,0$\\ \hline
        4 & $2,2,1,1,1,1,2,2,2,2,2,2,0$\\ \hline
        6 & $2,2,3,3,2,2,1,1,3,3,3,3,0$\\ \hline
        8 & $2,2,2,2,4,4,1,1,3,3,4,4,0$\\ \hline
        10 & $5,5,1,1,4,4,3,3,2,2,5,5,0$\\ \hline
        12 & $3,3,4,4,2,2,5,5,1,1,6,6,0$\\ \hline
        14 & $7,7,7,7,7,7,7,7,7,7,7,7,0$\\ \hline
        16 & $3,3,5,5,2,2,6,6,1,1,7,7,0$\\ \hline
        18 & $6,6,3,3,4,4,5,5,2,2,7,7,0$\\ \hline
        20 & $7,7,1,1,6,6,4,4,3,3,7,7,0$\\ \hline
        22 & $6,6,1,1,7,7,3,3,4,4,7,7,0$\\ \hline
        24 & $4,4,3,3,7,7,2,2,5,5,7,7,0$\\ \hline
        26 & $2,2,5,5,7,7,1,1,6,6,7,7,0$\\ \hline
    \end{tabular}
    \caption{$HH^d(\text{mf}(\C^5,\G,\check{W}))$ for $-10\leq d\leq 2$.}
    \label{tab:cE7hh}
\end{table}

\section{Verification from the physics side} \label{phy}

In this section, we substantiate our results from the physics perspective by invoking the AdS/CFT correspondence:
\begin{tcolorbox}
\begin{center}
large $N$ 4d $\cal N$=1 quiver SCFT 
 $\quad \longleftrightarrow\quad $ $K$-stable 3-fold singularity with NCCR.
\end{center}
\end{tcolorbox}\vspace{0.5em}
Crucially, the structure of the quiver and its superpotential is determined by the underlying NCCR. Throughout this work, we have confined ourselves to isolated terminal singularities, for which the existence of an NCCR is equivalent to that of a crepant resolution - thus yielding a transparent physical interpretation of the duality. Strong evidence for this AdS/CFT correspondence comes from matching invariant data on both sides, particularly the leading order central charge $a$ and the Hilbert series of the affine ring of X \cite{Xie:2019qmw,fang2023dimensionalquotientsingularity4d}\footnote{The chiral ring of the gauge theory splits into mesonic and baryonic sectors. The mesonic sector is captured by the coordinate ring of the symmetric product $M_{vac}=X^N/S_N$. Moreover, the mesonic operators splits into single-trace and multi-trace sectors. The single-trace mesonic operators are precisely those arising from a single copy of the affine coordinate ring $\mathbb{C}[X]$ of $X$. The full mesonic ring corresponds to the coordinate ring $\mathbb{C}[M_{vac}]=(\mathbb{C}[X]^{\otimes N})^{S_N}$ of the symmetric product $M_{vac}$, which is the algebra of symmetric functions on $\mathbb{C}[X]$. For further details, the readers can refer to \cite{Xie:2019qmw}.}.

On the field theory side, central charge $a$ of the SCFT can be computed from the quiver Hilbert series \cite{eager2014equivalence}. On the geometric side, $a$ is inversely proportional to the volume of the associated Sasaki–Einstein manifold, which itself can be computed from the singularity’s Hilbert series. Thus, a necessary condition for a consistent holographic duality is the equality of the two Hilbert series:
$$ H_{sing}(t)=H_{00}(t),$$
where $H_{sing}(t)$ denotes the Hilbert series of the singularity, and $H_{00}(t)$ the quiver Hilbert series at the distinguished node $0$.

Assuming the AdS/CFT correspondence, the existence of a crepant resolution can be tested by classifying all candidate quiver SCFTs - without prior knowledge of their superpotentials - whose quiver Hilbert series at a chosen node $0$ coincides with the Hilbert series of the singularity. Such gauge theories must also satisfy additional consistency requirements, which we detail in Subsection \ref{consquiv}. Our objective is to demonstrate that, for each singularity listed in Problem \ref{problem}, no admissible quiver SCFT exists. By AdS/CFT correspondence, this absence of field-theoretic candidates implies the nonexistence of crepant resolutions for these singularities. The explicit search procedure is described in Subsection \ref{implem}.

It is important to emphasize that this correspondence remains conjectural: matching Hilbert series provides only a necessary condition for holographic duality. Even if the series coincide, one must still compare additional data—such as the operator spectrum in the field theory versus those in the gravity theory—to confirm a genuine duality. However, for our “no-go” argument, verifying the mismatch of Hilbert series alone is sufficient.

\subsection{Consistency conditions for superconformal quiver gauge theory duals} \label{consquiv}

\paragraph{Scale invariance, unitarity and central charges.} A key requirement for consistency is the vanishing of the NSVZ $\beta$-functions for each gauge coupling \cite{Novikov:1983uc}. Equivalently, the ABJ anomaly for $U(1)_R$ should vanish
\begin{align}
    \text{Tr}(RG_iG_i)=0. \label{abjr}
\end{align}
Here the trace runs over all Weyl fermions charged under the simple gauge group $G_i$. In terms of group-theoretic data, this condition becomes
\begin{align}
    C_2(G_i)+\sum_{k \text{ chiral}}T(Rep_k)(R_k-1)=0, \label{beta}
\end{align}
where $C_2(G_i)$ is the quadratic Casimir of the adjoint representation of $G_i$ and $T(Rep_k)$ denotes the Dynkin index of the irreducible representation $Rep_k$, where the $k$-th chiral superfield transforms. The $R_k$ denotes the $R$ charge of the k-th chiral under $Rep_k$ representation of $G_i$. In our quiver theories, only adjoint and bifundamental chiral multiplets appear. For later reference, we summarize the Dynkin indices for the adjoint and fundamental representations of the $SU(N)$ gauge group:
$$T(adj)=N=C_2(SU(N)), \qquad T(fund)=\frac{1}{2}.$$

Another important constraint is unitarity, which requires every gauge-invariant chiral operator to have scaling dimension $\D\geq 1$. Using the 4d $\mathcal{N}=1$ shortening condition for chiral operators
$$ \Delta=\frac{3}{2}R,$$
the unitarity requires that $R\geq \frac{2}{3}$. Any gauge invariant chiral operator saturating this bound is free. Moreover, this shortening condition also ensures that $R$ charge of a composite chiral operator is simply the sum of the
$R$ charges of its constituent chiral fields.

The central charges $a$ and $c$ depend on the $U(1)_R$ symmetry via the 't Hooft anomalies $\Tr R$ and $\Tr R^3$ \cite{Anselmi:1997am}:
$$
a=\frac{3}{32}(3 \Tr R^3-\Tr R),\quad c=\frac{1}{32}(9 \Tr R^3-5\Tr R)\,.
$$
All such anomalies can be computed directly from the quiver data and the assumed $R$ charge assignments.
On the other hand, one may extract the leading large $N$ central charge from the asymptotic behavior of the quiver Hilbert series 
$H_{00}(t)$ \cite{Martelli:2005tp}. Writing $t=e^{-s}$, one finds
\begin{equation*}
H_{00}(\exp(-s))=\frac{a_0}{s^3}+\frac{a_1}{s^2}+\cdots\,.
\end{equation*}
Accordingly, the leading coefficient $a_0$ in the small-s expansion of the Hilbert series directly fixes the central charge $a$ and $c$ of the dual SCFT. \footnote{In the large $N$ limit, for quivers consisting of only adjoints and bifundamentals, $TrRGG=0$ ensures that $TrR=0$, therefore $a=c$.}
\begin{equation}\label{eqa}
a=c=\frac{27}{32}\frac{1}{a_0}N^2\,.
\end{equation}

\paragraph{Stability of chiral ring.}
Under the AdS/CFT dictionary, the requirement that a polarized chiral ring (chiral ring with specialized $R$ symmetry) is “stable” in the sense of four-dimensional
N=1 SCFTs is expectedly equivalent to the K-stability of the dual singularity \cite{collins2016k}. Concretely, one implements this stability by a generalized 
a-maximization procedure over trial U(1) mixings of the 
R-symmetry \cite{collins2016k,Fazzi_2020}. If the chiral ring fails this test—because a superpotential deformation becomes irrelevant or an operator’s 
R-charge falls below the unitarity bound—the putative SCFT does not exist, even when the geometry admits a noncommutative crepant resolution. Hence, imposing 
K-stability on the singularity provides a simple and powerful field-theoretic criterion to rule out such inconsistent quiver candidates. 

We therefore impose K-stability on the singularity, which restricts us to the region of parameter space listed in Table \ref{table:cdvlist} \cite{collins2016k,collins2019sasaki}. Concretely, this entails two conditions:
\begin{enumerate}
    \item \textbf{Positivity of Futaki invariants} Futaki invariants $F(X,\zeta,\eta)$ of some test configurations generated by $\eta$
are positive. Let $\zeta$ be the Reeb vector field encoding the $R$ charges $w_1,w_2,w_3.w_4$ of coordinates $x,y,z,w$  normalized s.t. the $(3,0)$ form $\Omega=\frac{dx\wedge dy \wedge dz\wedge dw}{df} $ has $R$ charge two, i.e., $w_1+w_2+w_3+w_4-d=2.$ For each test configuration generated by a vector field $eta$ with weight $(v_1,v_2,v_3,v_4)$, one computes the Futaki invariant $F(X,\zeta,\eta)$. Theorem 3.1 of \cite{collins2019sasaki} states that K-stability requires $$F(X,\zeta,\eta)>0,\,\text{for each nontrivial }\eta.$$
In our coordinate basis,
\begin{equation}
 \begin{aligned}
&F(X,\zeta,\eta)=-[v_4 w_1 w_2 w_3\left(w_1+w_2+w_3-2 w_4-d\right)\\
&+v_3 w_1 w_2 w_4\left(w_1+w_2+w_4-2 w_3-d\right)\\
&+v_2 w_1 w_3 w_4\left(w_1+w_3+w_4-2 w_2-d\right)\\
&+v_1 w_2 w_3 w_4\left(w_2+w_3+w_4-2 w_1-d\right)].
\end{aligned}
 \end{equation}

\item \textbf{Unitarity bound on $R$ charges} Since each coordinate $x,y,z,w$ correspond to a gauge-invariant operator under holographic dual. Unitarity requires the $R$ charges of them should be greater than or equal to $\frac{2}{3}$.
\end{enumerate}

These two criteria together carve out the K-stable region of $(N,k)$, as recorded in Table \ref{table:cdvlist}. We now illustrate the application of these two criteria with a specific example.

\textbf{Example:}
Consider the family of singularities $$X:\,\,x_1^2+x_2^2+x_3^4+z^k=0.$$ The $R$ charges of $(x_1,x_2,x_3,z;d)$ is $$(w_1,w_2,w_3,w_4;d)=(\frac{12k}{k+12},\frac{8k}{k+12},\frac{6k}{k+12},\frac{24}{k+12};\frac{24k}{k+12}).$$  To test K-stability, take the symmetry $\eta=(0,0,0,1)$. A straightforward computation gives the Futaki invariant $$F(X,\zeta,\eta)=-\frac{1152 (-24 + k) k^3}{(12 + k)^4}.$$
$F>0$ gives the constraints $0<k<24.$ The $k$ should be greater than 1 since it is an isolated singularity. Within the range $1<k<24$, the $R$ charges of $(x_1,x_2,x_3,z)$ are all larger than $\frac{2}{3}$, which satisfies the unitarity requiements.

\paragraph{NCCR and Shape of the quiver.}
Because the gauge–theory quiver is encoded by NCCR of the singularity, we must examine all admissible NCCRs. In fact, the quiver underlying any NCCR is directly related to the dual graph of a corresponding crepant resolution. Morrison \cite{Morrison1985} proved that any small crepant resolution of an isolated Gorenstein threefold singularity yields a collection of exceptional curves whose intersection graphs are those shown in Figure \ref{fignccr}.

\begin{figure}[h]
    \centering
  \begin{tikzcd}[column sep=small]
  1\ar[r,bend left=15]& 2\ar[l,bend left=15]\ar[r,bend left=15]\ar[loop,out=290,in=250,looseness=6]\ar[loop,out=120,in=80,looseness=4]&3\ar[l,bend left=15]\ar[r,bend left=15]\ar[loop,dashed,out=290,in=250,looseness=6]&\cdots\ar[l,bend left=15]\ar[r,bend left=15]\ar[loop,dashed,out=290,in=250,looseness=6]&n\ar[l,bend left=15]\ar[loop,dashed,out=290,in=250,looseness=6]&\overline{E}_n, n=3,4,5 
  \end{tikzcd}
\begin{tikzcd}[column sep=small,row sep=small]
    &&1\ar[d,bend left=15]\ar[dl,bend left =15]\\
    2\ar[r,bend left=15]\ar[loop,dashed,out=290,in=250,looseness=6]&3\ar[l,bend left=15]\ar[r,bend left=15]\ar[ur,bend left=15]&4\ar[l,bend left=15]\ar[loop,dashed,out=290,in=250,looseness=6]\ar[u,bend left=15]\ar[r,bend left=15]&5\ar[l,bend left=15]\ar[r,bend left=15]\ar[loop,dashed,out=290,in=250,looseness=6]&\cdots\ar[l,bend left=15]\ar[r,bend left=15]\ar[loop,dashed,out=290,in=250,looseness=6]&n\ar[l,bend left=15]\ar[loop,dashed,out=290,in=250,looseness=6]& \Tilde{E}_n,n=6,7,8
\end{tikzcd}
\begin{tikzcd}[column sep=small, row sep=small]
     1\ar[r,bend left=15]\ar[loop,out=290,in=250,looseness=6]\ar[loop,out=120,in=80,looseness=4]&\cdots\ar[l,bend left=15]\ar[r,bend left=15]\ar[loop,dashed,out=290,in=250,looseness=6]&n\ar[l,bend left=15]\ar[loop,dashed,out=290,in=250,looseness=6]&\overline{D}_n,n\geq 1
\end{tikzcd}
\begin{tikzcd}[column sep=small,row sep=small]
    &1\ar[d,bend left=15]\ar[loop,dashed,out=120,in=80,looseness=4]\\
2\ar[loop,dashed,out=290,in=250,looseness=6]\ar[r,bend left=15]&3\ar[l,bend left=15]\ar[r,bend left=15]\ar[u,bend left=15]\ar[loop,dashed,out=290,in=250,looseness=6]&4\ar[l,bend left=15]\ar[r,bend left=15]\ar[loop,dashed,out=290,in=250,looseness=6]&\cdots\ar[l,bend left=15]\ar[r,bend left=15]\ar[loop,dashed,out=290,in=250,looseness=6]&n\ar[l,bend left=15]\ar[loop,dashed,out=290,in=250,looseness=6]&D_{n},n\geq 4
\end{tikzcd}
\begin{tikzcd}[column sep=small,row sep=small]
    1\ar[r,bend left=15]\ar[loop,dashed,out=290,in=250,looseness=6]&\cdots\ar[l,bend left=15]\ar[r,bend left=15]\ar[loop,dashed,out=290,in=250,looseness=6]&n\ar[l,bend left=15]\ar[loop,dashed,out=290,in=250,looseness=6]&A_{n},n\geq 1
\end{tikzcd}
\begin{tikzcd}[column sep=small,row sep=small]
&&1\ar[loop,dashed,out=120,in=80,looseness=4]\ar[d,bend left=15]\\
2\ar[r,bend left=15]\ar[loop,dashed,out=290,in=250,looseness=6]&3\ar[l,bend left=15]\ar[r,bend left=15]\ar[loop,dashed,out=290,in=250,looseness=6]&4\ar[l,bend left=15]\ar[r,bend left=15]\ar[loop,dashed,out=290,in=250,looseness=6]\ar[u,bend left=15]&5\ar[l,bend left=15]\ar[r,bend left=15]\ar[loop,dashed,out=290,in=250,looseness=6]&6\ar[l,bend left=15]\ar[r,bend left=15]\ar[loop,dashed,out=290,in=250,looseness=6]&\cdots\ar[l,bend left=15]\ar[r,bend left=15]\ar[loop,dashed,out=290,in=250,looseness=6]&n\ar[l,bend left=15]\ar[loop,dashed,out=290,in=250,looseness=6]&E_n,n=5,6,7
\end{tikzcd}
\begin{tikzcd}[column sep=small,row sep=small]
&1\ar[dl,bend left=15]\ar[d,bend left=15]\\
2\ar[loop,dashed,out=290,in=250,looseness=6]\ar[r,bend left=15]\ar[ur,bend left=15]&3\ar[l,bend left=15]\ar[r,bend left=15]\ar[loop,dashed,out=290,in=250,looseness=6]\ar[u,bend left=15]&\cdots\ar[l,bend left=15]\ar[r,bend left=15]\ar[loop,dashed,out=290,in=250,looseness=6]&n\ar[l,bend left=15]\ar[loop,dashed,out=290,in=250,looseness=6]&\Tilde{D}_n,n\geq 3
\end{tikzcd}
    \caption{The possible shape of the quivers corresponding to the (one-node-deleted) NCCR of the cDV singularities. \cite{august2020tilting} The dotted arc represents that there may be an adjoint chiral or may be not.}
    \label{fignccr}
\end{figure}
More precisely, the diagrams in Figure \ref{fignccr} depict the quivers associated to NCCR with the distinguished node deleted\footnote{The deleted distinguished node corresponds to the coordinate ring it self as its module. It appears as a summand in any tilting module whose endomorphism algebra produces an NCCR.}. Each remaining node (labelled by its multiplicity $1,2,\cdots,n$) represents an exceptional $\mathbb{P}^1$ in the small resolution, and each pair of opposite arrows encodes transverse intersection between two such curves.

To reconstruct the 4d $\mathcal{N}=1$ gauge theory quiver \cite{decelis2015flops}, one proceeds as follows:\begin{enumerate}
    \item \textbf{Gauge node} Associate to each node as an $SU$ type gauge group whose rank is proportional to $N$, the number of $D3$ branes probing the singularity.
    \item \textbf{Distinguished node} Add a node corresponding to the trivial module.
    \item \textbf{Bifundamental fields} Associate a pair of bifundamental chiral multiplets for each pair of opposite arrows between two nodes.
    \item \textbf{Adjoint fields}  A solid loop at a node indicates the presence of an adjoint chiral multiplet transforming in the adjoint of that gauge group; dashed loops may or may not correspond to additional adjoint fields.
\end{enumerate}
When a singularity is known to admit crepant resolution, one can often construct the corresponding NCCR explicitly—for example, by employing matrix factorization methods. In contrast, for singularities whose crepant resolvability remains undecided (as in our case), a brute‐force search for all candidate quiver SCFTs requires enumerations. Morrison’s classification of small crepant resolutions and their associated dual graphs enables us to restrict to a finite set of quiver topologies, avoiding unnecessary enumeration.
 
Since we only focus on the $f(X)=1,2$ case, i.e., the associated quiver have two-node or three-node. Among the quivers shown in Figure \ref{fignccr}, only the $A_n$ and $\overline D_n$ families admit two and three-node quivers. Therefore, most general two-node and three-node quivers are those shown in Figure \ref{fig23node}. We emphasize that these general quivers also includes quivers with fewer adjoint fields: setting the $R$-charge of an adjoint chiral to $1$ effectively removes its contribution in the infrared, recovering the cases with less adjoint at that node.

\begin{figure}[h]
    \centering
 \begin{subfigure}[b]{0.45\textwidth}
 \centering
     \begin{tikzcd}
	N_1\ar[loop,out =290,in=250,looseness=9,"a_2"]\ar[loop,out =110,in=70,looseness=9,"a_1"]\ar[r,bend left=10,"c_1"]&N_2 \ar[l,bend left=10,"d_1"]\ar[loop,out =290,in=250,looseness=9,"b_2"]\ar[loop,out =110,in=70,looseness=9,"b_1"]
\end{tikzcd}
\caption{quiver for one mass deformation} \label{figtwonode}\end{subfigure}
    \begin{subfigure}[b]{0.45\textwidth}
    \centering
 \begin{tikzcd}   &N_2\ar[loop,out=110,in=70,looseness=9,"b_1"]\ar[dr,bend left=10,""]\ar[dl,bend left=10]&\\
N_1\ar[loop,out=290,in=250,looseness=9,"a_2"]\ar[loop,out=110,in=70,looseness=9,"a_1"]\ar[ur,bend left=10,""]\ar[rr,bend left=10,""]&&N_3\ar[loop,out =290,in=250,looseness=9,"c_2"]\ar[loop,out =110,in=70,looseness=9,"c_1"]
  \ar[ul,bend left=10,""]\ar[ll,bend left=10,""]\end{tikzcd}\caption{quiver for two mass deformations}
  \label{figthreenode}
 \end{subfigure}
    \caption{There are at most two adjoints on the distinguished node $N_1$. There may be a pair of bifundamental chirals between all the other nodes and the distinguished node. }
    \label{fig23node}
\end{figure}

 For the case $f=1$, the quiver consists of two nodes associated to gauge groups $G=SU(N_1),SU(N_2)$. Imposing the NSVZ beta functions \cite{Novikov:1983uc} for each gauge group give equations 

\begin{equation}
 \left\{   \begin{aligned}
&N_1(1+R_{a_1}-1+R_{a_2}-1)+\frac{1}{2}(R_{c_1}-1+R_{d_1}-1)N_2=0,\\
&N_2(1+R_{b_1}-1+R_{b_2}-1)+\frac{1}{2}(R_{c_1}-1+R_{d_1}-1)N_1=0.\\
    \end{aligned}\right.
\end{equation}
Here $R_{X}$ denote $R$ charges of the corresponding chiral field $X=a_1,a_2,b_1,b_2,c_1,d_1$. Physical consistency further requires both ranks to be positive: $N_1>0,N_2>0.$ One can solve the ratio of ranks $N_1,N_2$ of the gauge groups from the beta functions in terms of $R$ charges of the quiver. 

For a given quiver gauge theory $Q$,  one computes its matrix Hilbert series \cite{bocklandt2008graded,fang2023dimensionalquotientsingularity4d} via the formula 
\begin{equation}\label{eqMQ}
H(Q,t)=\frac{1}{1-M_Q(t)+t^2M_Q^T(t^{-1})-t^2}\,,
\end{equation}
where adjacent matrix $M_Q$ can be read from the quiver and $R$ charges:

\begin{enumerate}
\item If $i\neq j$, the off-diagonal element $M_{ij}$ of $M_Q$ is  $$M_{ij}=\sum_{\text{bifund chirals in} (\mathbb{N}_i,\mathbb{\bar{N}}_j)}t^{R_{ij}}\,.$$
\item If $i=j$, the diagonal element $M_{ii}$ is 
$$
M_{ii}=\sum_{\text{adjoint chiral fields}}t^{R_{ii}}\,.
$$
\end{enumerate}
Here $R_{ij}$ and $R_{ii}$ denote the $R$ charge of  bifundamental and adjoint chiral fields, respectively. The $(i,j)$ entry of $H(Q,t)$ counts the oriented path from the node $i$ to node $j$ with the $R$ charge grading.
In particular, the quiver Hilbert series $H_{00}(Q,t)$ with respect to the distinguished node $0$ is the $(0,0)$ entry of the matrix $H(Q,t)$, which enumerates closed loops based at node, and therefore counts the gauge invariant scalar operators. Under the holographic duality, the $H_{00}$ is believed to be identified with the Hilbert series of the affine coordinate ring of the dual singularity.
Note that each node can be chosen as the node $0$ and the corresponding quiver Hilbert series may be different. Therefore, in our search procedure, we compute and compare $H_{00}$ for every possible node acting as node 0.

Now let us give an example of K-stable cDV singularity which is known to have CRs, along with the field theory dual. 
\vspace{1.0em} 

\textbf{Morrison-Pinkham example} 

Consider the two-node quiver $Q$:
\begin{tikzcd}
	N\ar[r,bend left=10,"\frac{1}{2}"]&2N \ar[l,bend left=10,"\frac{1}{2}"]\ar[loop,out =290,in=250,looseness=9,"\frac{1}{2}"]\ar[loop,out =110,in=70,looseness=9,"\frac{3}{4}"]\,
\end{tikzcd}.

The $R$ charges of bifundamental chiral fields $c$ and $d$ are $\frac{1}{2}$, while the two adjoint of the right node have
$R_{b_1}=\frac{3}{4},~ R_{b_2}=\frac{1}{2}$. Therefore, the matrix $M_Q$ is 
$$
M_Q=\begin{pmatrix}
   t^{R_{a_1}}+t^{R_{a_2}}& t^{R_{c_1} }\\
   t^{R_{d_1}}& t^{R_{b_1}}+t^{R_{b_2}}
\end{pmatrix}=\begin{pmatrix}
  0& t^{\frac{1}{2}}\\
   t^{\frac{1}{2}}& t^{\frac{1}{2}+\frac{3}{4}}
\end{pmatrix}.
$$
With the left node as the distinguished node, the quiver Hilbert series can be derived from $M_Q$ using formula (\ref{eqMQ})
\begin{equation*}
H_{00}=   \frac{1-t^{\frac{18}{4}}}{\left(1-t^{\frac{9}{4}}\right) \left(1-t^{\frac{6}{4}}\right) \left(1-t^{\frac{4}{4}}\right) \left(1-t^{\frac{7}{4}}\right)}\,.
\end{equation*}
This indicates that the corresponding geometry is generated by four fields of weights
$$(x,y,z,w)=(\frac{9}{4}, \frac{6}{4}, \frac{4}{4}, \frac{7}{4}),$$ 
and that the degree $\frac{18}{4}$ relation is
\begin{equation*}
x^2+y^3+yz^3+w^2z=0\,.
\end{equation*}
This is precisely the $cD_4$ Morrison-Pinkham example, known to admit an NCCR \cite{Aspinwall_2012}. 

\subsection{Implementation of the search}  \label{implem}
We now outline our algorithm for identifying candidate quiver theories whose Hilbert series 
$H_{00}(t)$ matches that of each singularity in Problem \ref{problem}. Although a large number of quivers might a priori satisfy the Hilbert series condition, scale invariance, unitarity, and the allowed quiver topologies (determined by $f(X)+1$ nodes) reduce the search space dramatically. For simplicity, we impose three additional constraints: 

\begin{enumerate}
\item All $R$ charges of the quiver are in the range $[0,2]$.\footnote{ There are two reasons of the simplification: 1) outside this range, there are some terms of the Hilbert series has negative powers of $t$. Therefore, the Hilbert series is  hard to coincide those from the singularity side.
2) the quiver with $R$ charges outside this range is much more difficult to find a sensible superpotential.\cite{Bajc:2019vbp}}
    \item Each pair of bifundamental chiral fields is assigned the same $R$ charge due to symmetry.
    \item If the common denominator $R$ charges of the variable $x,y,z,w$ is $m$, then we restrict all $R$ charges of quiver to lie in the discrete set  $$\{0,\frac{1}{m},\frac{2}{m},\cdots,\frac{2m}{m}\},$$ 
    since $x,y,z,w$ are composites of the fundamental chiral fields. 
\end{enumerate}

With these constraints, we can exhaustively search all possible quivers with one, two or three nodes. As a consistency check, our program reproduces Morrison-Pinckham example described above.\footnote{One can see the attached \texttt{Mathematica} notebook\texttt{ Find dual.nb}.} We then verify that no further $N=1$ superconformal quiver exists, with the prescribed number of nodes, for the remaining cases:
$$\boxed{
\begin{aligned}
&J=E_6: \quad x^2+y^3+z^4+y t^k: \Big\{\makecell{ k\in\{2,3,5,6,7,9,10,11,13,14,15\} \text{  (2-node quivers)} \\ k\in\{4,12\} \text{  (3-node quivers)}} \\
& J=E_7: \quad x^2+y^3+y z^3+z t^{k}: k\in\{4,6,8,10,12,16,18,20,22,24,26\} \text{  (2-node quivers)}\end{aligned}}$$
In each instance, no admissible quiver with a correct Hilbert series is found, confirming the absence of a superconformal dual. This absence verifies our main claim.

\section{Conclusions and future directions} \label{con}
In this paper we established our main claim through the novel mirror-symmetry based approach. It gives a new way to understand the existence of 4d $\mathcal N=1$ superconformal quiver gauge theory duals of all singularities in Table \ref{table:cdvlist}, with $J=E_6,E_7$ and $E_8$. Most of these singularities fall into the $cE_n$ types, with the notable exception of the classical $cD_4$ Morrison-Pinkham example. We verify this implication from the physics side, by searching for all possible field theory candidates. Although the search is subject to additional constraints, it provides compelling evidence for the claim. Finally, we conclude with a list of open questions which worth further investigation. 

\paragraph{More general singularities}

One may consider more general classes of singularities. The most immediate extensions are canonical singularities, and beyond them, Kawamata log-terminal (klt) singularities. Unlike cDV singularities, divisors may appear in a resolution of a canonical singularity. From a physics viewpoint one might imagine engineering gauge theories by wrapping D7-branes on such resolved geometries. However, to the best of our knowledge, no systematic study in the literature carries out this generalization. In particular, the relation between NCCRs and crepant resolutions is no longer transparent, and it is unclear whether the quivers with potentials produced from NCCRs should still be interpreted as the genuine quiver gauge theories and their superpotentials. For klt singularities, the situation is even less understood.

Within this broader framework, it is also natural to include non-isolated singularities, whose links are in general not smooth. Examples of this type already appear in \cite{Fazzi_2020}\footnote{The non-isolated singularities considered in this paper admit at least two $\C^\ast$ actions, which in particular allows for irrational $R$ charges.}. On the algebraic side, one may still construct noncommutative crepant resolutions via matrix factorizations. However, the non-isolated nature of singular locus means that any crepant resolution must be contract divisors, and the connection between such divisorial resolutions and the associated NCCR quivers is once again unclear. If the NCCRs do correspond to physical theories, the resulting quiver gauge models may involve chiral multiplets with irrational 
$R$ charges (and hence irrational scaling dimensions).

It is also natural to investigate quotient singularities. Most cyclic quotients have been analyzed in the literature. As shown in \cite{chen20174dn2scftsingularity}, except for product of cyclic quotients, the singularities are typically non-isolated. In particular, quotients of $\C^3$ by finite subgroups of $SO(3)$, as studied in \cite{decelis2015flops,fang2023dimensionalquotientsingularity4d}, are generically non-isolated. Nevertheless, the proposed dual field‑theory candidates in these works remain highly compelling.

Another direction is to search for new dual pairs of cDV singularities beyond the list of Wang and Xie. Once the Gorenstein condition is relaxed, three-fold terminal singularities are not necessarily hypersurfaces, and may instead arise as complete intersections or through other constructions. Exploring such cases may lead to previously unknown gauge/gravity duals. 

\paragraph{Dualities between superconformal quiver gauge theories}
In carrying out the search described in Section \ref{phy}, we discovered that the same algorithm also identifies candidate dualities among superconformal quiver gauge theories. Specifically, different quiver realizations sometimes produce the same Hilbert series. To further test whether these theories are truly dual, one can compare their large $N$ superconformal indices (or single‑trace indices) \cite{Gadde:2010en} and central charges. However, because matching these invariants is only a necessary—not a sufficient—condition for duality, each pair remains a provisional candidate that merits additional checks.
The gauge groups should be viewed as $U$ type instead of $SU$ type. They may have the same holographic dual. These dualities are worth studying, as they do not belong to either the classic Seiberg duality framework \cite{Seiberg:1994pq} or its recent extensions \cite{Bajc:2019vbp, Fang:2024nqy}.
Unfortunately, these theories are not holographic dual to any singularity considered in this paper.  

\begin{figure}[H]
    \centering
\begin{subfigure}[b]{0.45\textwidth}\centering
\begin{tikzcd}   2N \ar[r,bend left=10," \frac{2}{3}"]&
6N \ar[loop,out=110,in=70,looseness=9," \frac{4}{9}"]\ar[l,bend left=10,""]\ar[r,bend left=10, " \frac{1}{3}"]&3N\ar[l,bend left=10,""]\ar[loop,out=110,in=70,looseness=9,"\frac{4}{3}"]\end{tikzcd}  \caption{ }\end{subfigure}
\begin{subfigure}[b]{0.45\textwidth}
\centering\begin{tikzcd}   2N \ar[loop,out=110,in=70,looseness=9," \frac{4}{3}"]\ar[r,bend left=10," \frac{1}{3}"]&
4N\ar[loop,out=110,in=70,looseness=9," \frac{4}{3}"]\ar[l,bend left=10,""]\ar[r,bend left=10, " \frac{1}{3}"]&6N\ar[l,bend left=10,""]\ar[loop,out=110,in=70,looseness=9,"\frac{4}{9}"]\end{tikzcd}\caption{ }\end{subfigure}
    \caption{Two quiver gauge theories in which every gauge node is an unitary gauge group. Both theories have $TrR=0$. $TrR^3 $ are all $\frac{2560 N^2}{81}$. Furthermore, we have checked that their superconformal indices coincide in the large $N$ limit.}
    \label{fig:enter-label}
\end{figure}

In fact, additional duality candidates can be uncovered by first matching quiver Hilbert series and then verifying other physical data—such as central charges and large $N$ superconformal indices. Note that the quiver Hilbert series need not be tied to the particular singularities studied here, this approach therefore provides a general algorithm for discovering new duality candidates.

\acknowledgments
We thank Dan Xie for providing the foundational insights and background research that inspired this work. ZY would like to thank Fulin Xu for various informative and inspiring conversations on birational geometry, Kazushi Ueda and Kenji Fukaya for some points on symplectomorphisms. We would also like to thank Michael Wemyss and Ban Lin for discussions. Finally, we express our gratitude to Robert Valandro for correspondences.

\bibliographystyle{JHEP}
\bibliography{main}
\end{document}